\documentclass[twocolumn]{aastex631}
\usepackage{graphicx}
\usepackage{amsmath}

\usepackage{subfigure}
\usepackage{cleveref}
\let\bfseries\mdseries

\graphicspath{{figures/}}

\begin{document}

\title{Galaxy Model Subtraction with a Convolutional Denoising Autoencoder}

\author[0000-0003-0685-3525]{Rongrong Liu}
\affiliation{Center for Astrophysics $\vert$ Harvard \& Smithsonian, 60 Garden Street, Cambridge, MA 02138, USA}

\author[0000-0002-2073-2781]{Eric W. Peng}
\affiliation{NSF's NOIRLab, 950 N.\ Cherry Avenue, Tucson, AZ 85719, USA}

\author[0000-0003-4672-8497]{Kaixiang Wang}
\affiliation{Department of Astronomy, Peking University, Beijing 100871, People’s Republic of China}
\affiliation{Kavli Institute for Astronomy and Astrophysics, Peking University, Beijing 100871, People’s Republic of China}

\author[0000-0002-8224-1128]{Laura Ferrarese}
\affiliation{Herzberg Astronomy and Astrophysics Research Centre, National Research Council of Canada, Victoria, BC V9E 2E7, Canada}

\author[0000-0003-1184-8114]{Patrick C\^{o}t\'{e}}
\affiliation{Herzberg Astronomy and Astrophysics Research Centre, National Research Council of Canada, Victoria, BC V9E 2E7, Canada}

\begin{abstract}
Galaxy model subtraction removes the smooth light of nearby galaxies so that fainter sources (e.g., stars, star clusters, background galaxies) can be identified and measured. Traditional approaches (isophotal or parametric fitting) are semi-automated and can be challenging for large data sets. We build a convolutional denoising autoencoder (DAE) for galaxy model subtraction: images are compressed to a latent representation and reconstructed to yield the smooth galaxy, suppressing other objects. The DAE is trained on GALFIT-generated model galaxies injected into real sky backgrounds and tested on real images from the Next Generation Virgo Cluster Survey (NGVS). \textbf{ To quantify performance, we conduct an injection–recovery experiment on residual images by adding mock globular clusters (GCs) with known fluxes and positions. Our tests confirm a higher recovery rate of mock GCs near galaxy centers for complex morphologies, while matching ellipse fitting for smooth ellipticals. Overall, the DAE achieves subtraction equivalent to isophotal ellipse fitting for regular ellipticals and superior results for galaxies with high ellipticities or spiral features. } Photometry of small-scale sources on DAE residuals is consistent with that on ellipse-subtracted residuals. Once trained, the DAE processes an image cutout in $\lesssim 0.1$ s, enabling fast, fully automatic analysis of large data sets. We make our code available for download and use.

\end{abstract}

\keywords{}

\section{Introduction} 

Galaxy model subtraction is a process used in astronomical image analysis to isolate celestial objects within an image by removing the contribution of the galaxy itself. This process is often needed to identify and measure the photometry of celestial objects projected close to the centers of nearby galaxies, such as globular clusters \citep{Blakeslee_1995ApJ, Jordan_2009ApJS} or gravitationally lensed background sources \citep[e.g.,][]{Sonnenfeld_2020A&A}. For example, in \cite{Blakeslee_1995ApJ}, the residual images from galaxy model subtraction are used to measure the surface brightness fluctuations due to globular clusters and calculate the specific frequencies and luminosity function widths of globular clusters. 
Successful galaxy model subtractions can allow researchers to study a broader range of objects in the Universe with better accuracy. 

A simple approach to removing smooth galaxy light is with a ring median filter \citep{Secker_1995}. It is easy to implement and relatively fast. This filter estimates variations in the image by calculating median intensities within a circular annulus around each pixel. In principle, if there is a source at the center, then as long as the inner radius of the annulus is larger than the spatial scale of the object, the object should not bias the estimate of the local background, which can subsequently be subtracted. The ring median filter can subtract features on scales larger than the ring diameter and works well for large elliptical galaxies with smoothly-varying brightness profiles. However, the ring median filter is unable to remove features that are smaller than its inner radius, which at minimum is a set to be a few times the width of the point spread function. Also, when ring sizes are small, the possibility of a single source in the ring biasing the local measurement becomes higher. This filtering technique therefore suffers from over-subtraction near sources and fails to accurately model the steep change in brightness profiles at the centers of a galaxies. This is especially problematic for lower-mass galaxies, which are smaller in size, and whose effective radii can approach the seeing scale in ground-based imaging. For example, dwarf galaxies in the nearby Virgo galaxy cluster have $R_e\approx 10\arcsec$ \citep{Ferrarese_2006}.

Among the most commonly used techniques to perform galaxy model subtraction is ellipse fitting \citep{Jedrzejewski_1987, Tonry_1997ApJ}, which fits elliptical isophotes to a galaxy's light at different radii and iteratively adjusts the parameters (like ellipticity and position angle) to minimize the intensity variations along each isophote. The ellipse fitting typically involves some manual efforts in fitting and subtracting galaxy models from the observed images, and is therefore difficult to apply to large datasets. Also, the results produced by ellipse fitting on non-elliptical galaxies are usually less accurate since those galaxies have uneven luminosity profiles and complicated internal structures \citep{Peng_2010}. 

\begin{figure*}
\centering
\subfigure{ 
    \includegraphics[width=0.30\textwidth]{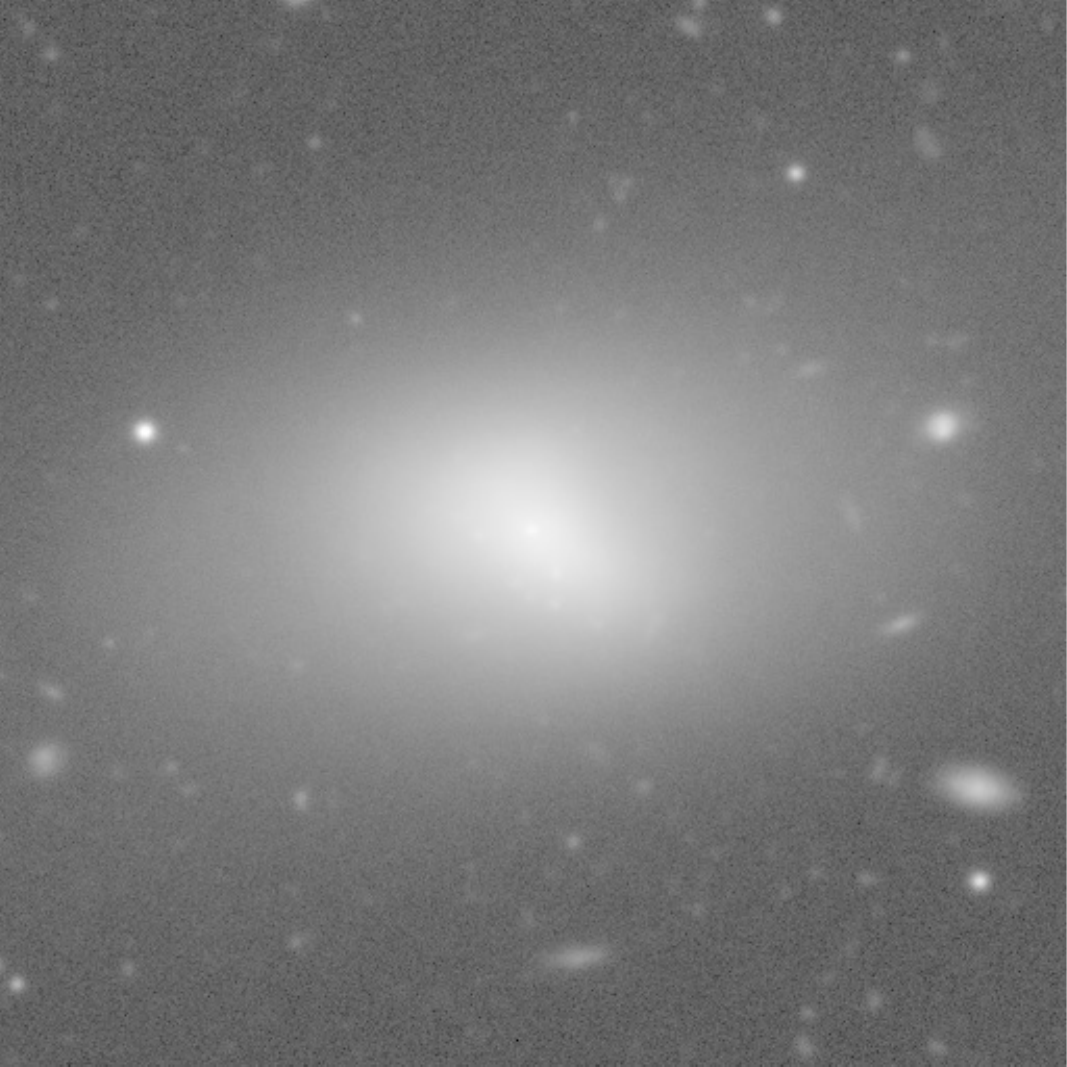}
}
\hfill
\subfigure{ 
    \includegraphics[width=0.30\textwidth]{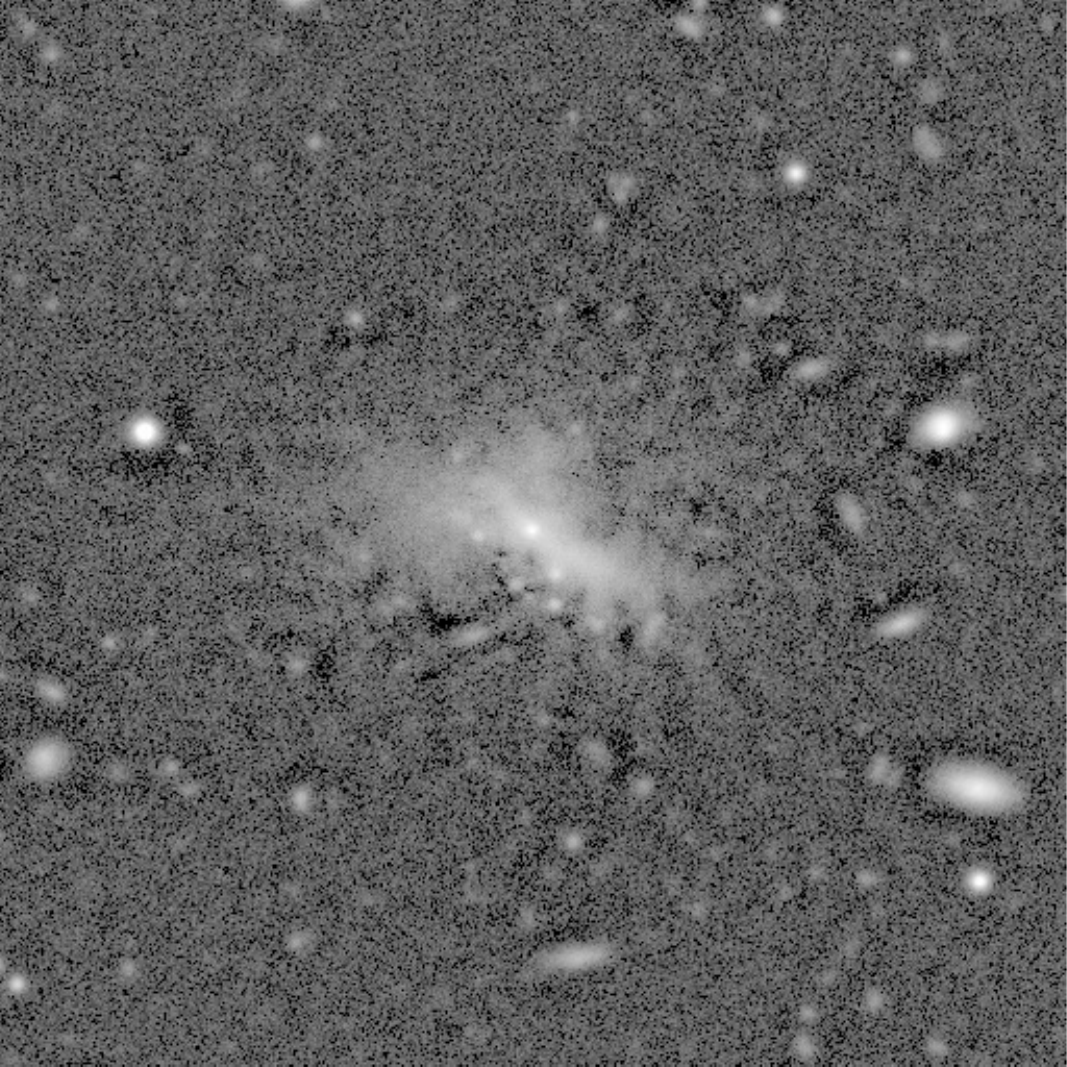}
}
\hfill
\subfigure{ 
    \includegraphics[width=0.30\textwidth]{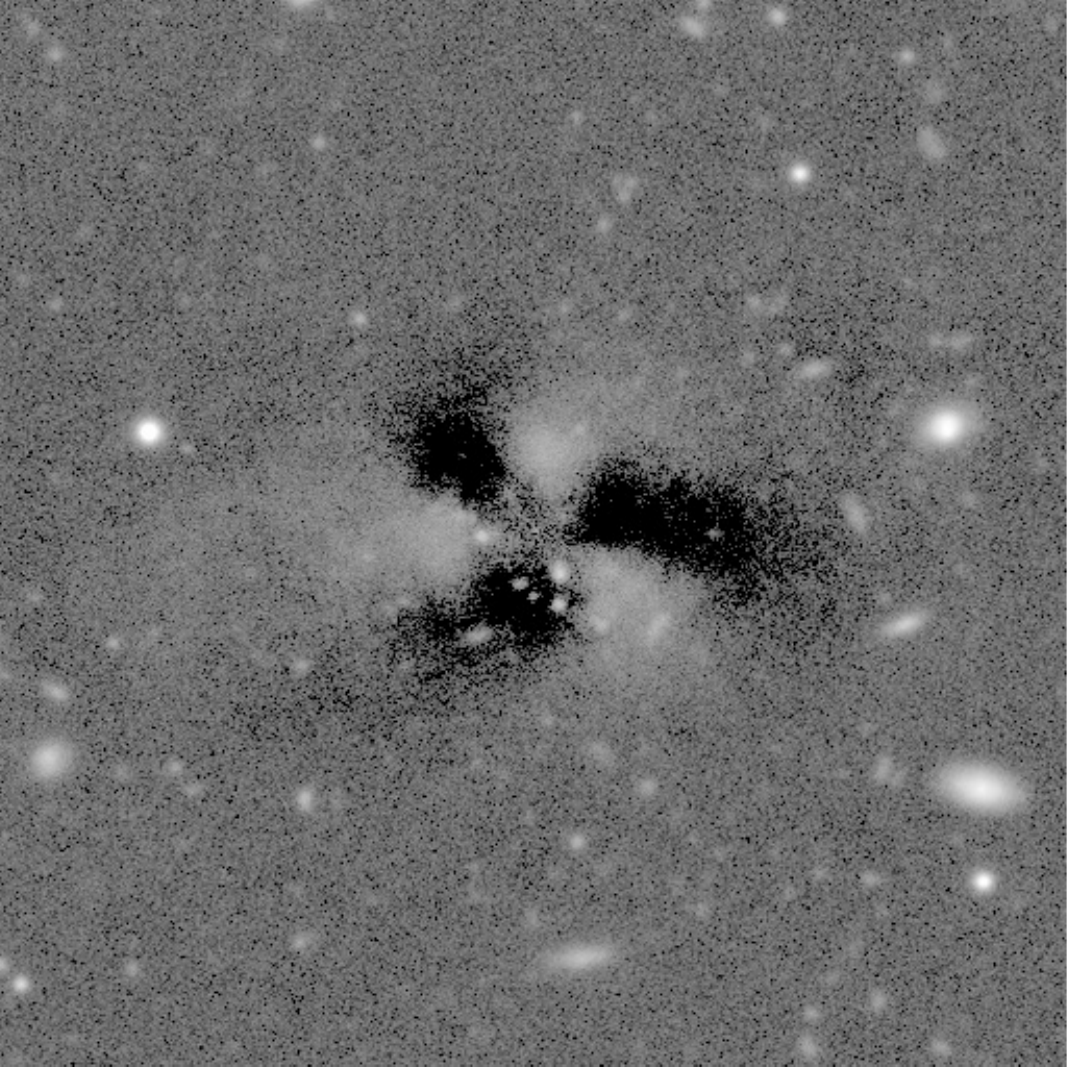}
}
\caption{Galaxy image and model subtraction results for VCC 0407 using traditional methods. The left image shows the original $g'$-band image of VCC 0407, an early-type dwarf galaxy in the Virgo cluster with $M_{g'}=-16.82$~mag. The size of this image is $512\times 512$ pixels, corresponding to $95\farcs7\times 95\farcs7$ in observed units. The center image shows the residual image produced by a ring median filter with an inner radius of 11 pixels (2\farcs057) and an outer radius of 16 pixels (2\farcs992). The right image shows the residual image produced by ellipse subtraction. The ring median filter is unable to subtract the bright galaxy center, while the ellipse fitting struggles to fit the spiral structures in this galaxy.}
\label{fig:trad_methods}
\end{figure*}

\textbf{Figure~\ref{fig:trad_methods} shows an example of galaxy model subtraction on VCC 0407, a Virgo Cluster galaxy, using a ring median filter (middle) and ellipse fitting (right). From the residual image of ring median subtraction, we can see that it leaves dark rings around objects due to over-subtraction and cannot subtract the central bright region of the galaxy. On the other hand, the ellipse fitting struggles with features like a bar and spiral arms in the galaxy since it's designed to fit only elliptical isophotes to the galaxy. These issues with traditional galaxy subtraction methods make it challenging to distinguish objects projected near the galaxy center from the light of the galaxy itself.}

To overcome these challenges, we employ machine learning methods to develop a more efficient and accurate technique for galaxy model subtraction. Machine learning is a category of fast-developing data analysis techniques that train the models to process data in desired ways without a fixed algorithm. Machine learning methods can be divided into two major subcategories: supervised learning and unsupervised learning. In supervised learning, a model is trained on a pre-labeled dataset to do tasks such as classification and regression. In unsupervised learning, the dataset comes unlabeled, and a model is trained to recognize the underlying structure of the data (\citealt{murphy_2012}). Deep learning is a subset of machine learning where more than three neural network layers (input layer, output layer, at least one hidden layer) are present in the model. \textbf{Depending on the task and available data, deep learning models can be trained in either supervised or unsupervised settings.  Deep learning has proven to be an extremely versatile tool in astronomy, where it has been used for galaxy image classification (e.g. \citealt{a&a_2020_638, Gharat_2022}), time-domain event detection (e.g. \citealt{Muthukrishna_2022, Villar_2021}), and spectral analysis (e.g. \citealt{10.1093/mnras/stx3298})}


Convolutional neural networks (CNNs) are a type of deep learning model well suited for image processing tasks. \textbf{The convolutional layers consists of a set of kernels that move across the input image dimensions to create a feature map representing the location and intensity of features in the image. Each kernel focuses on a small (typically $3\times 3$ or $5\times 5$ pixels) region in the input data. The structure of CNNs allows them to learn features at different spatial scales. As the network goes deeper, the convolutional layers will build on the simple features to learn more complicated ones.  The shallower layers of the CNNs can learn small features like the smoothness of the local brightness profile, while deeper layers can capture larger features like the overall morphology of the galaxy. } 

An autoencoder is a type of unsupervised deep learning model designed to learn a low-dimensional representation (the ``latent space'') from image data and reconstruct the original input from the learned representation. \textbf{ When dealing with astronomical dataset, unsupervised learning is often preferred because the data is often unlabeled. Autoencoders have been widely applied in galaxy modeling, for example in galaxy morphology classification and image generation \citep{Spindler_2020, Smith_2023}, and in inferring physical parameters from the latent space \citep{Aragon-Calvo}}

The denoising autoencoder (DAE) is a specialized type of autoencoders introduced by \citet{VincentLBM08} that takes corrupted data as input and learns to reconstruct the clean data. This training process allows DAEs to learn noise-invariant features in data. Since then, the DAE has been applied in diverse disciplines like medical image analysis \citep{Gondara_2016_Med} and speech enhancement \citep{Yu_2020}. It has also achieved success in astronomy, such as denoising radio astronomical images to detect faint radio sources \citep{Gheller_2021} and denoising optical spectra of galaxies \citep{10.1093/mnras/stad2709}. These successful applications of denoising autoencoders across multiple domains suggest its potential in galaxy model subtraction.  

In this paper, we propose the usage of a Convolutional Denoising Autoencoder (DAE) for galaxy model subtraction.  In our model, the DAE keeps the light from the central galaxy and remove the sky noise, stars, and background galaxies from the image. Subtracting the reconstructed model galaxy from the input image gives a clean residual image.

In the following sections, we will introduce the model architecture of the DAE and implementation details, and present galaxy model subtraction results obtained from the DAE with comparisons to results from traditional methods.

\section{Data}
At a distance of 16.5~Mpc \citep{Blakeslee_2009}, the Virgo cluster is the nearest galaxy cluster, but is at a distance where galaxies are not resolved into stars in typical ground-based imaging. The corresponding distance modulus is $m-M = 31.09$~mag, which we use throughout this work to convert between apparent and absolute magnitudes. Although the need for fast modeling of smooth galaxies extends to much larger distances, Virgo is an ideal place to explore the use of new methods due to the availability of the Next Generation Virgo Cluster Survey (NGVS; \citealt{ngvs}). The NGVS is a comprehensive optical imaging survey of the Virgo cluster conducted using the  1 deg$^2$ MegaCam instrument on Canada-France-Hawaii Telescope. The survey covers a total area of 104 deg$^2$ in the $u^*g'i'z'$ bandpasses and reaches a point-source depth of $g' \approx 25.9$~mag and a surface brightness limit of $\mu_{g'} \sim 29$~mag~arcsec$^{-2}$. The NGVS provides high-quality and deep data of nearby galaxies, similar to next-generation surveys like Rubin/LSST. Therefore, developing a galaxy modeling technique that works on the NGVS data will allow applications on future datasets from upcoming large surveys. All the work in this paper is done with optical $g'$-band images, although the work could easily be extended to other filters.

\begin{figure*}[]
    \centering
    \subfigure{ 
        \includegraphics[width=0.45\textwidth]{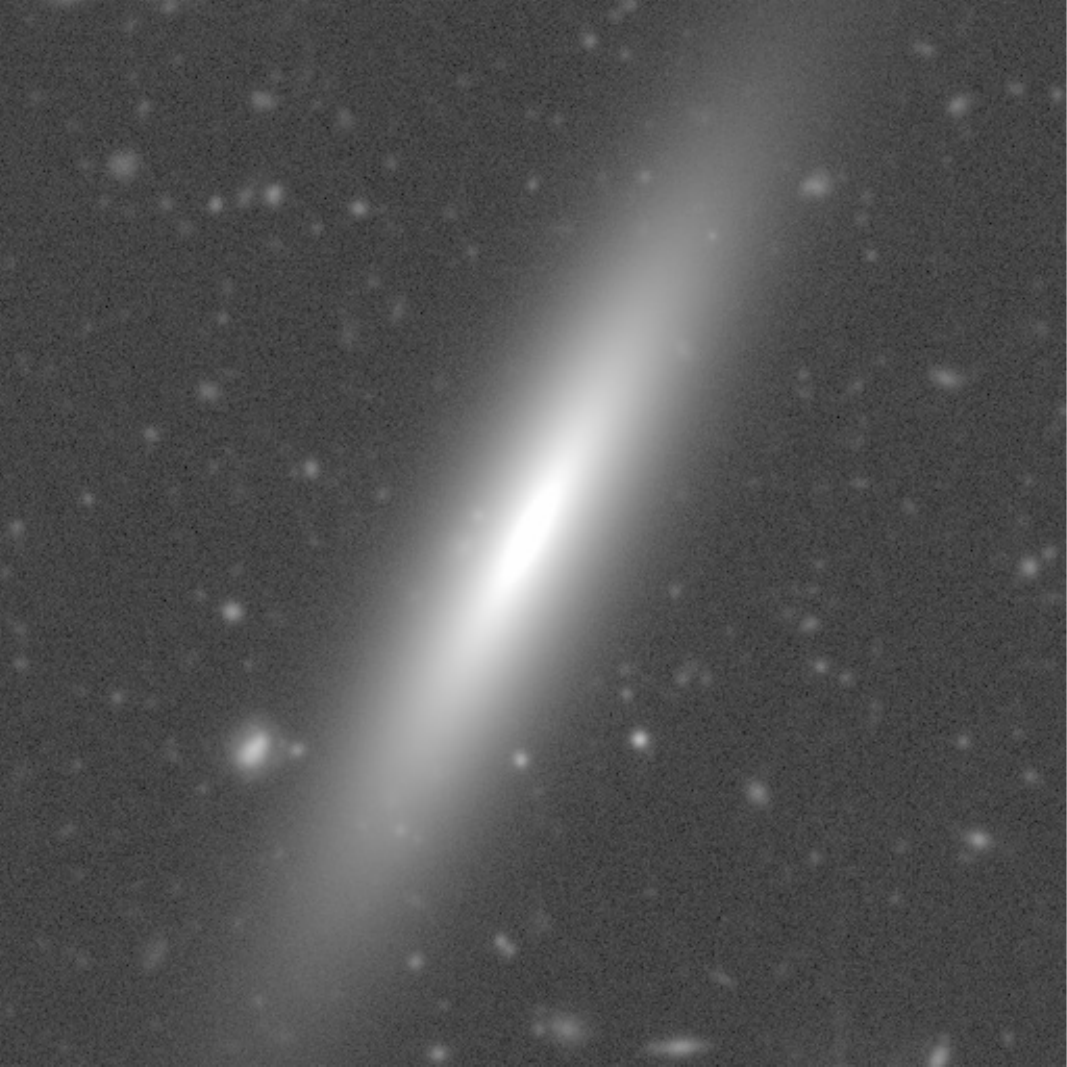}
    }
    \hfill
    \subfigure{ 
        \includegraphics[width=0.45\textwidth]{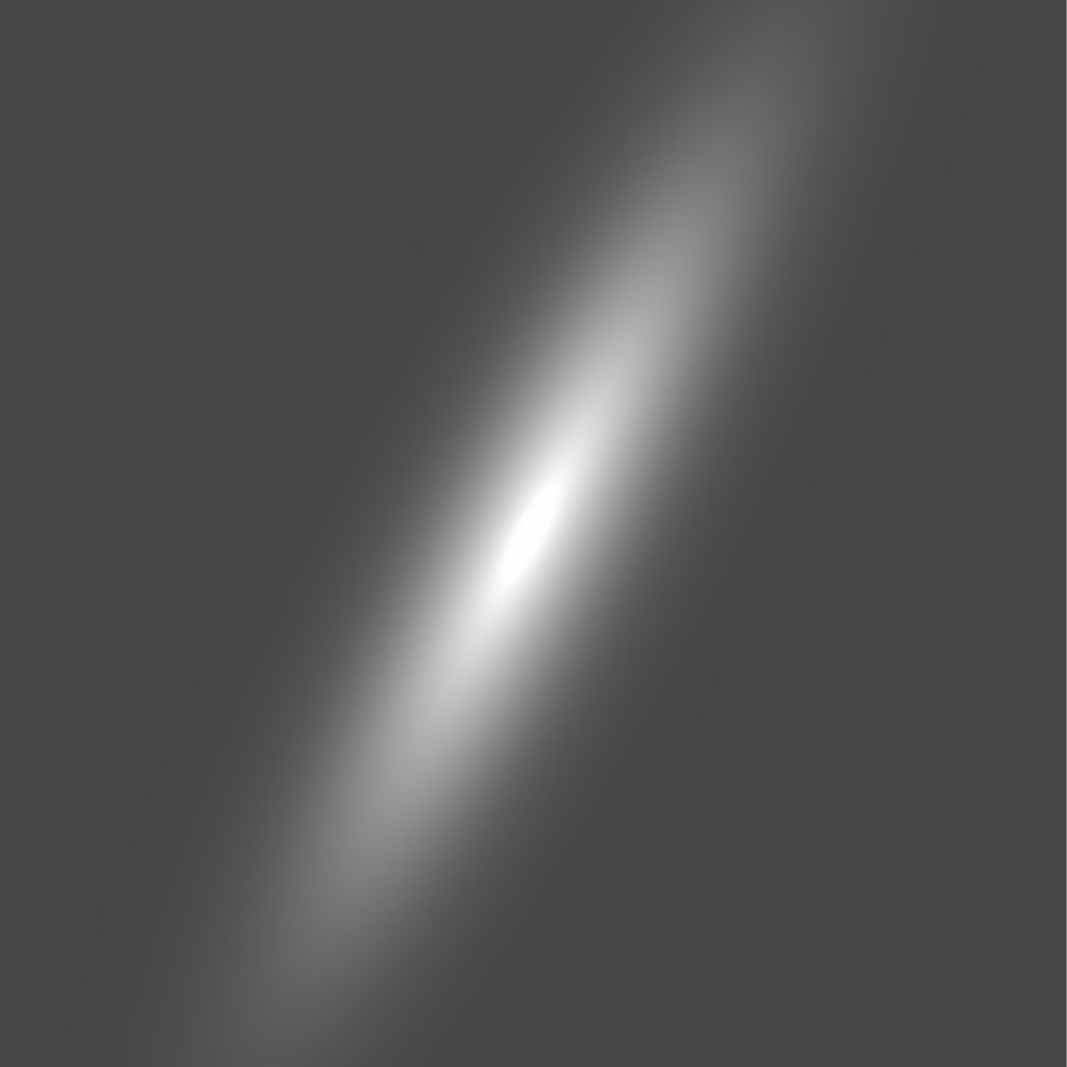}
    }
    \caption{An example pair of images in the training set. We use the injected galaxy image (left) as input and the clean model galaxy image (right) as targeted output. Each image is of size $512\times 512$ pixels ($95.7\times 95.7$ arcsecs in observed units). The left image is the injected galaxy image (or ``noisy" image in training) produced by injecting GALFIT galaxy models into ``blank sky'' background images taken from NGVS dataset. The sky background images are chosen so they do not contain  bright stars or large galaxies. The right image is the clean model galaxy image (or ``clean'' image in training) generated using GALFIT package. We bin the training data by apparent magnitudes. This model galaxy ($g'=14.41$~mag) was used in the training set for galaxies with $14<g'<15$~mag.}
    \label{fig:training_pair}
\end{figure*}

The dataset used to train our DAE consists of two parts: the galaxy images injected with sky background and the clean model galaxy images. An example pair of images in the dataset is shown in Figure~\ref{fig:training_pair}. In order to produce realistic galaxies for this training set, we used GALFIT \citep{Peng_2010} to produce model galaxies that follow a \citet{Sersic_1968} profile. The S\'ersic model parameters used to generate model galaxies---the S\'ersic index $n$,  the effective radius $R_e$, and the effective surface brightness $\mu_e$---are chosen randomly from a Gaussian distribution around the mean relations for these quantities for Virgo cluster early-type galaxies. We use the same relations as presented in \citet{Lim_2020}. 

The sky background images are taken from random coordinates in the NGVS dataset. After retrieving the sky background images, we manually cleaned the dataset by removing images with large galaxies and bright stars. We then add the sky background images to clean model galaxy images produced by GALFIT, thus creating an injected galaxy image corresponding to each clean model galaxy. 

The galaxy images are placed into different bins of apparent magnitudes ($[12,13), [13,14), [14,15)$), and one model is trained for each of the dataset bin. We are interested in galaxies in the observed magnitude range $12<g'<15$~mag because our main goal for performing galaxy model subtraction is to study the globular clusters within galaxies. For larger galaxies, the distribution of globular clusters are more extended compared to the galaxy light, but in these dwarf galaxies, the globular clusters are spatially distributed with half-number radii that are comparable to the galaxy's size. Therefore, it is important to model the galaxy light more accurately in this magnitude range. We bin the galaxy images by integrated magnitude because, according to the size-luminosity relation, the galaxies in the same magnitude bin will have similar sizes and surface brightness profiles. Binning allows us to optimize the model parameters effectively for different scales. 


The complete dataset contains 12000 pairs of images in each integrated magnitude bin. The size of the images is $512\times 512$ pixels, or $95\farcs7\times 95\farcs7$. 

\section{Methods}

\subsection{Denoising Autoencoder}

We use a Convolutional DAE to separate clean galaxy from observed galaxy images. An autoencoder is a type of neural network that can extract essential features of the input data and reconstruct the data based on the encoded representation. 
The autoencoder has two main components: an encoder and a decoder. The encoder takes the input data and compresses it into a latent space representation that has lower dimension than the input layer. If the autoencoder is properly trained, its latent space will contain the key features from the original input data (i.e., the smooth galaxy). The decoder takes the encoded representation and reconstructs the data such that the output resembles the targeted output. 
Mathematically, a general autoencoder is defined on input/output vector space $\mathbb{R}^n$ and the encoded vector space $\mathbb{R}^p$, which are real vector spaces with dimensionality $n, p \in \mathbb{Z^+}$, the set of positive integers. For any pair of functions $A: \mathbb{R}^p\rightarrow\mathbb{R}^n$ and $B:\mathbb{R}^n\rightarrow\mathbb{R}^p$, the autoencoder can transform an input vector $\mathbf{x}$ into an output vector $A(B(\mathbf{x}))$. 
Given a set of $m$ training vectors $\{\mathbf{x}_1, ..., \mathbf{x}_m\}\in \mathbb{R}^n$, a set of $m$ corresponding target vectors $\{\mathbf{y}_1,...,\mathbf{y}_m\}\in \mathbb{R}^n$, and a loss function $\Delta$, an autoencoder aims to learn functions $A$ and $B$ such that the overall loss function 
\begin{equation}
    \sum_{t=1}^m \Delta \big(A(B(x_t)), y_t\big)
    \label{eqn:loss}
\end{equation}
is minimized \citep{ae_math}, where $t$ is the index for input vector. The loss function $\Delta$ is a measure of difference between the output vector and the target vector. 



A DAE takes noisy data as input  $\mathbf{x}$  and learns to produce clean (denoised) data as output $\mathbf{y}$, removing noise in the process. In our study, we inject model galaxy images into real blank sky observations as the input $\mathbf{x}$ and clean model galaxy images (with no ``sky'' background) as the targeted output $\mathbf{y}$.


\begin{figure*}[ht]
    \centering
\includegraphics[width = \linewidth]{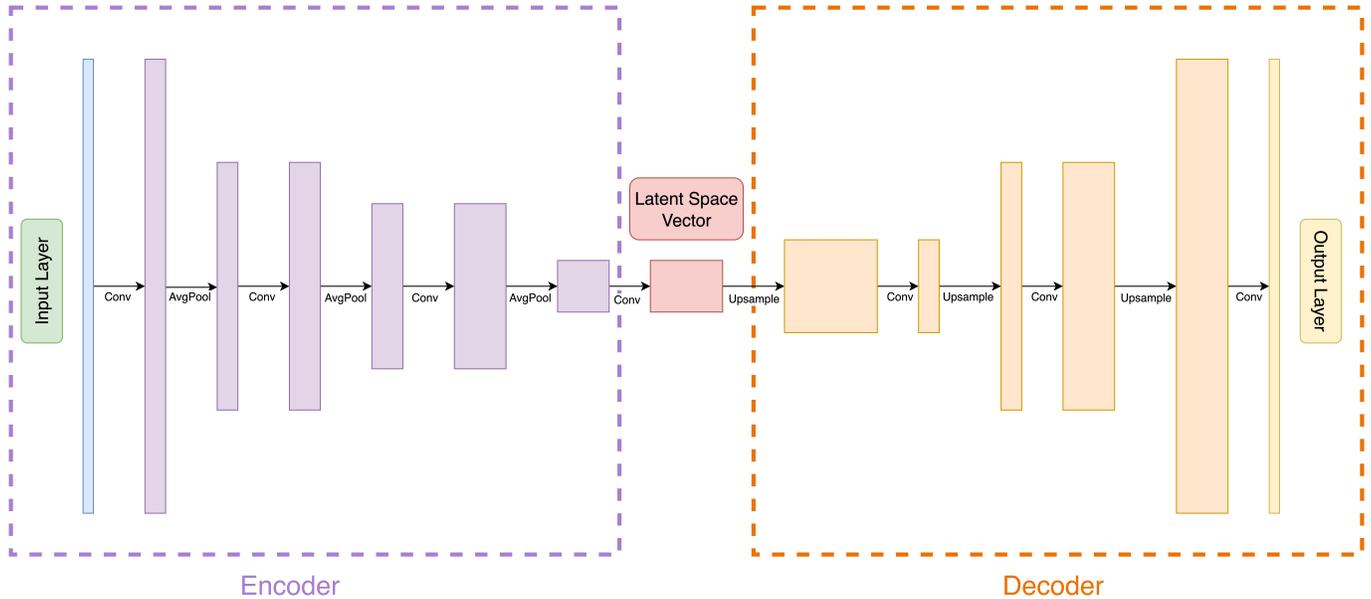}
        \caption{Graphical illustration of our DAE model architecture. The encoder (left)  compresses the input data into a lower-dimensional latent space vector with key features.  The decoder (right) reconstructs the input data from this latent space into a target output which shares some key features with input data but is not identical to the input data. In our denoising autoencoder, the input data are a group of observed galaxy images (with the galaxy, foreground stars, and background objects), and the output data are the corresponding clean galaxy-only images.}
    \label{fig:autoencoder_architecture}
\end{figure*}

We build the denoising autoencoder with \texttt{TensorFlow.keras} \citep{tensorflow2015-whitepaper,chollet2015keras} package in Python.  Our model architecture consists of three encoder blocks and three decoder blocks. Each encoder block contains a convolutional layer followed by an average pooling layer, while each decoder block contains a convolutional layer followed by an upsampling layer. A graphical representation of our model architecture is shown in Figure \ref{fig:autoencoder_architecture}. We use a leaky ReLU (Rectified Linear Unit) activation function with slope = 0.1 after each convolutional layer. Activation functions add non-linearity to the model so it can fit the complex patterns in the data better. A standard ReLU activation $f(x) = \max(0,x)$ sets all negative inputs to 0, while the leaky ReLU allows a small, non-zero gradient for negative inputs ($f(x) = \max(ax,x)$, $a=0.1$ in our setup) to optimize the gradient flow during training.  The filter size, the pooling function, and the activation function are determined by choosing the best combination from experiments. The shapes of the input and output of the autoencoder are both $512\times 512\times 1$. In total, there are 109,001 trainable parameters in the model. The detailed architecture of the autoencoder is shown in Table~\ref{tab:architecture}. 

\begin{table*}
\begin{tabular}{cccccc}
\hline
Layer \# & Layer Type          & Input Shape            & Filter Size        & Output Shape              & \# of Parameters \\ \hline
1            & Convolution             & $512\times 512\times 1$ & $3 \times 3 \times 16$ & $512\times 512\times 16$ & 160                  \\
2            & AveragePooling     & $512\times 512\times 16$ & $2 \times 2$       & $256\times 256\times 16$ & 0                    \\
3            & Convolution             & $256\times 256\times 16$ & $3 \times 3 \times 32$ & $256\times 256\times 32$ & 4,640                 \\
4            & AveragePooling     & $256\times 256\times 32$ & $2 \times 2$       & $128\times 128\times 32$ & 0                    \\
5            & Convolution             & $128\times 128\times 32$ & $3 \times 3 \times 64$ & $128\times 128\times 64$ & 18,496                \\
6            & AveragePooling     & $128\times 128\times 64$ & $2 \times 2$       & $64\times 64\times 64$   & 0                    \\
7            & Convolution             & $64\times 64\times 64$   & $3 \times 3 \times 128$ & $64\times 64\times 128$ & 73,856                \\
8            & UpSampling        & $64\times 64\times 128$  & $2 \times 2$       & $128\times 128\times 128$ & 0                    \\
9            & Convolution             & $128\times 128\times 128$ & $3 \times 3 \times 8$ & $128\times 128\times 8$ & 9,224                 \\
10           & UpSampling        & $128\times 128\times 8$ & $2 \times 2$       & $256\times 256\times 8$ & 0                     \\
11           & Convolution             & $256\times 256\times 8$ & $3 \times 3 \times 32$ & $256\times 256\times 32$ & 2,336                \\
12           & UpSampling        & $256\times 256\times 32$ & $2 \times 2$       & $512\times 512\times 32$ & 0                    \\
13           & Convolution             & $512\times 512\times 32$ & $3 \times 3 \times 1$  & $512\times 512\times 1$ & 289                  \\
\hline
 & & & &  Total & 109,001\\
\hline
\end{tabular}
\caption{The architecture of the convolutional image-denoising autoencoder used in tests}
\label{tab:architecture}
\end{table*}

\begin{figure}[]
    \vspace{1cm}
    \centering
    \includegraphics[width = \linewidth]{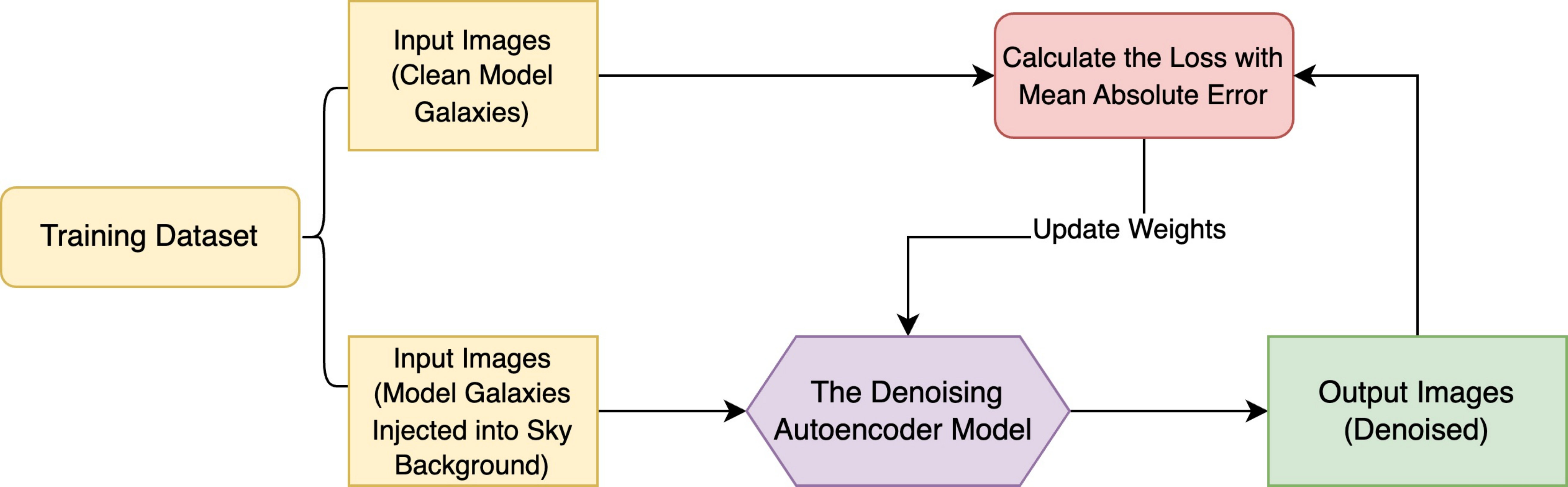}
    \caption{Training loop of the denoising autoencoder. During training, the ``noisy'' images (galaxies injected into sky backgrounds) are put through the denoising autoencoder, producing output denoised images. The output images are then compared with the clean model galaxy image to calculate the loss by Mean Absolute Error (MAE). Weights in the denoising autoencoder are updated by back propagation to minimize the loss. } 
    \label{fig:training_loop}
\end{figure}

During the training process, we optimize the denosing autoencoder's ability to separate the light of central galaxy from the image with sky background. As described in the data section, the training dataset comprises pairs of images: ``noisy'' input images containing model galaxies injected into real sky backgrounds, and their corresponding target images of clean model galaxies. The dataset is split into the training set (80\%) and validation set (20\%). For testing, we use real galaxy images from the NGVS dataset within the trained magnitude range. 

The model processes data in batches of 16 images. This batch size gives both high computational efficiency and stable gradient estimates. Each training iteration consists of the following steps: 
\begin{enumerate}
    \item The model takes a batch of noisy training images and reconstructs the denoised output images.
    \item The reconstruction error between the reconstructed images and their corresponding clean images is calculated using Mean Absolute Error (MAE) as the evaluation metric. MAE is defined as the average error in magnitude between predicted and actual values in each pixel of the image.
    \item The model uses back-propagation to calculate how each parameter contributes to the final reconstruction error and determines how to adjust the parameters (weights and bias) to reduce the error. 
    \item The parameters are updated using Adam (Adaptive moment estimation, \citealt{kingma2017adam}) as the optimizer. 
\end{enumerate}
The above training loop is repeated until the reconstruction error of the validation set converges. The model with the lowest validation MAE is chosen as the final model. Upon convergence, the model will have the ability to reconstruct denoised images that are close to clean galaxy images. The training is done on Google Colab\footnote{https://colab.research.google.com/} using an NVIDIA V100 GPU. With a dataset of 12000 images and a batch size of 16, the model typically takes approximately 3--4 hours to converge.  A schematic diagram of the training procedure is shown in Figure~\ref{fig:training_loop}. 

\subsection{Alternative Models}
In addition to the baseline DAE, we also explore some alternative models that can potentially improve the DAE's performance in handling image features such as bright stars and image artifacts. 

\subsubsection{Partial Convolution}
One technique we investigate is partial convolution \citep{liu2018imageinpaintingirregularholes}, an adaptive convolution method that processes images based on valid pixel masks. In partial convolution, the model updates the convolution filter based on the availability (defined by a binary mask) of data in each region, ensuring only valid pixels are accounted for in the output. This technique is usually used to treat irregular holes in the image by reconstructing a complete and contextually coherent image. In our implementation, we produce the binary masks using Source Extractor (SExtractor, \citealt{1996A&AS..117..393B}) to exclude pixels that contains bright objects on the residual images. 

In testing, we find that the training time for partial convolution is much longer compared to the baseline DAE model, taking more than 30 hours with the same computational resources (the baseline model takes 3--4 hours to converge). Also, even after substantial training time,  the loss fluctuates without converging to a low value. The reconstructed image quality did not show significant improvement compared to the standard denoising autoencoder. This is potentially due to the increase in model complexity and instabilities in gradient computation with masked regions. Based on these results, the current implementation of partial convolution requires further modifications in either its masking methods or optimization algorithm to be a useful extension of galaxy image denoising. 

\subsubsection{Variational Autoencoder}
Another model we test is a variational denoising autoencoder to increase the generalization capability of our model. Instead of directly encoding the image data into a compressed representation, VAEs encode data into a probabilistic distribution in the latent space. In our context, this means that a VAE can generate a more diverse representation of galaxy models and avoid overfitting the training dataset. A VAE can help the model deal with different galaxy morphologies that cannot be directly generated as training data using GALFIT. 

However, our experiments show that the VAE produces lower quality reconstructions compared to our standard denoising autoencoder. This is because the probabilistic nature of the latent space introduces additional uncertainty in the reconstruction process, resulting in less precise galaxy model recovery and increased blur in the output images. VAE produces over-subtractions near the center of the galaxies that can be subtracted easily by the standard denoising autoencoder. For our galaxy modeling task, the deterministic approach of standard autoencoders provides better performance than the more complex probabilistic framework of VAEs.

\section{Results}
\subsection{Residual Comparison}
In this section, we present a comparison between the results generated by the ellipse subtraction and the DAE subtraction on real galaxy images in NGVS. 

\textbf{The ellipse fitting results are produced by the same pipeline as described in Section 5.1 of \citet{Ferrarese_2020}. First, a mask is created for contaminants (e.g. foreground stars, background galaxies, globular clusters) in the image. Isophotes are then fitted to the target galaxy with the \texttt{ELLIPSE} task in \texttt{IRAF} (Image Reduction and Analysis Facility, \citealt{1986SPIE..627..733T,1993ASPC...52..173T}), following the algorithm of \citet{Jedrzejewski_1987} and \citet{1990AJ....100.1091P}. In this process, the surface brightness within an elliptical annulus is expanded into a Fourier series, and the parameters of the isophote (center, ellipticity,  position angle) are adjusted to minimize the residuals.  The masking and isophotal fitting steps are repeated to accurately exclude contaminants near the galaxy center. Finally, a smooth model galaxy is generated using the best-fitting parameters and subtracted from the original image to give a residual image. }

Figure \ref{fig:all_examples_14_15} shows the results on real galaxy images from the NGVS dataset with $14<g'< 15$~mag, cropped to a size of $512\times 512$~pixels. Figure \ref{fig:vcc_13to14_examples} shows the same results but on galaxies with $13<g'<14$~mag. 

For face-on dwarf elliptical galaxies (e.g., VCC~0033 in Figure \hyperref[fig:5a]{5(a)}), both the autoencoder and ellipse fitting effectively subtract the galaxy light and produce clean residuals. This is expected since the profile of a face-on elliptical galaxy matches well with the assumptions of ellipse fitting. 

Edge-on galaxies (e.g. VCC 1304 in Figure \hyperref[fig:5b]{5(b)}) are hard to model since their shapes are elongated and their light profiles change quickly near the center. It is difficult to fit elliptical isophotes to edge-on galaxies because there is often the presence of a disk. When ellipse fitting is used on these edge-on galaxies, it creates an X-shaped artifact in the residual. By contrast, the autoencoder directly extracts the distribution of galaxy light without predefined geometric assumptions and produces more uniform residual images. 

Furthermore, the denoising autoencoder can handle galaxy images with complex structures better than ellipse fitting. For example, VCC~0407 in Figure \hyperref[fig:5c]{5(c)} has spiral arm features where ellipse fitting leaves artifacts in the residual images (as also shown previously in Figure~\ref{fig:trad_methods}). These over- and under-subtraction patterns in the residuals can affect both the number of sources detected in the residual image and the photometric measurements of the sources. The autoencoder can subtract the galaxy light evenly and produce cleaner residuals without artifacts. 

\textbf{However, one limitation is that our current DAE model does not perform as well on galaxies with bulges. Because our DAE is only trained on single-S\'ersic models, it struggles to fit multiple components and leaves residuals in the inner regions (e.g. VCC 0575 in Figure \hyperref[fig:6c]{6(c)}). Ellipse fitting, which explicitly allows the isophotal parameters to vary with radius, can sometimes capture such multi-component profiles more successfully. However, future training of the DAE with multi-component model galaxies should improve the DAE's performance.}

In summary, the visual residual comparison shows that the denoising autoencoder is a more effective method for galaxy model subtraction compared to ellipse fitting. The autoencoder adapts well to different types of galaxies, from face-on elliptical galaxies to more complex spiral and edge-on galaxies, while ellipse fitting struggles with features like spiral arms and steep brightness changes in those galaxies. The autoencoder produces residuals with fewer artifacts and preserves the faint sources, making it easier to detect and measure objects like globular clusters.


\begingroup
\setlength{\textfloatsep}{0pt} 

\begin{figure*}
\centering

\subfigure{\parbox[b]{0.03\textwidth}{\raggedleft(a)}} 
\label{fig:5a}
\hfill
\subfigure{\parbox[b]{0.30\textwidth}{\centering
    \text{Original Galaxy Image}\\
    \includegraphics[width=0.29\textwidth]{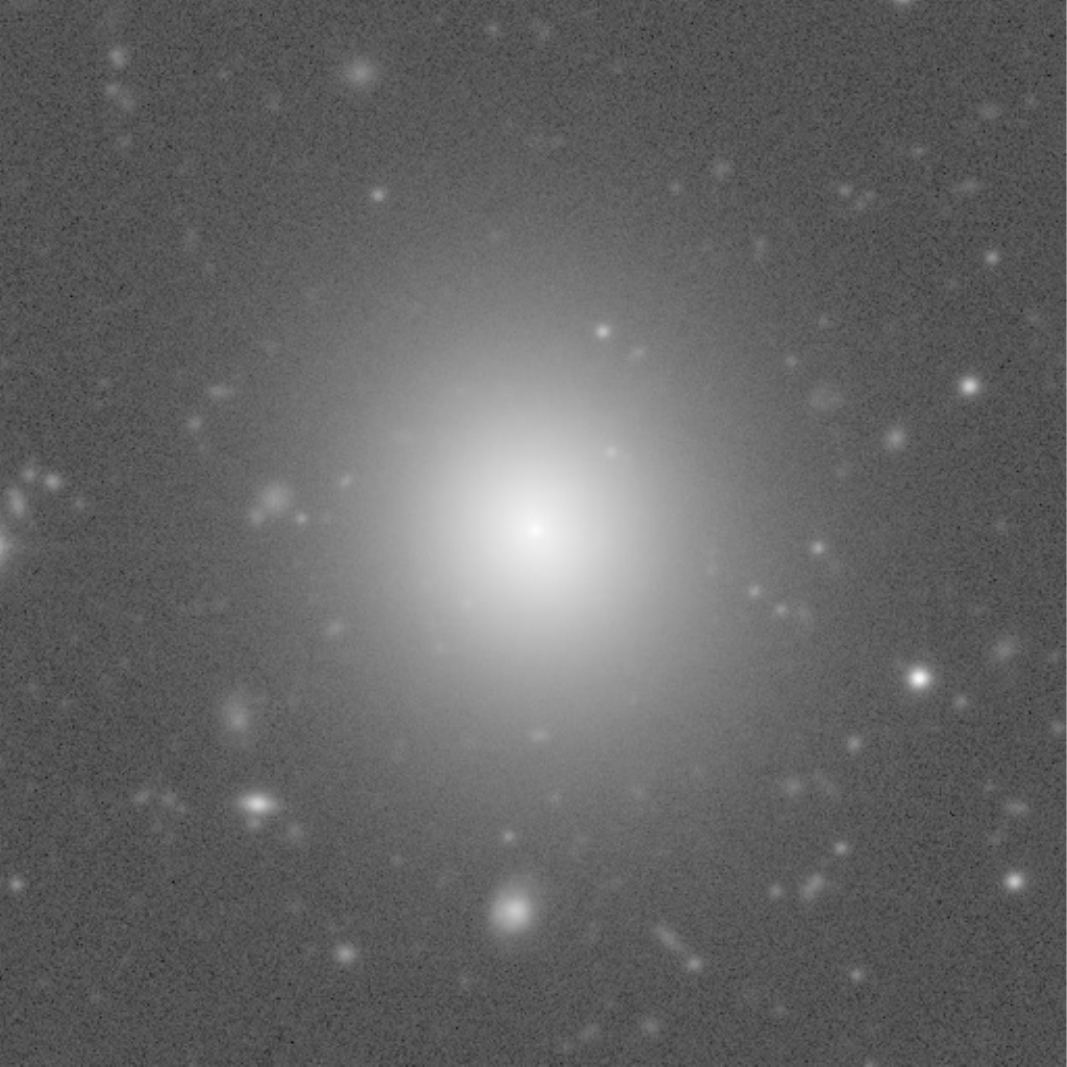}}}
\hfill
\subfigure{\parbox[b]{0.30\textwidth}{\centering
    \text{Ellipse Subtraction Residual}\\
    \includegraphics[width=0.29\textwidth]{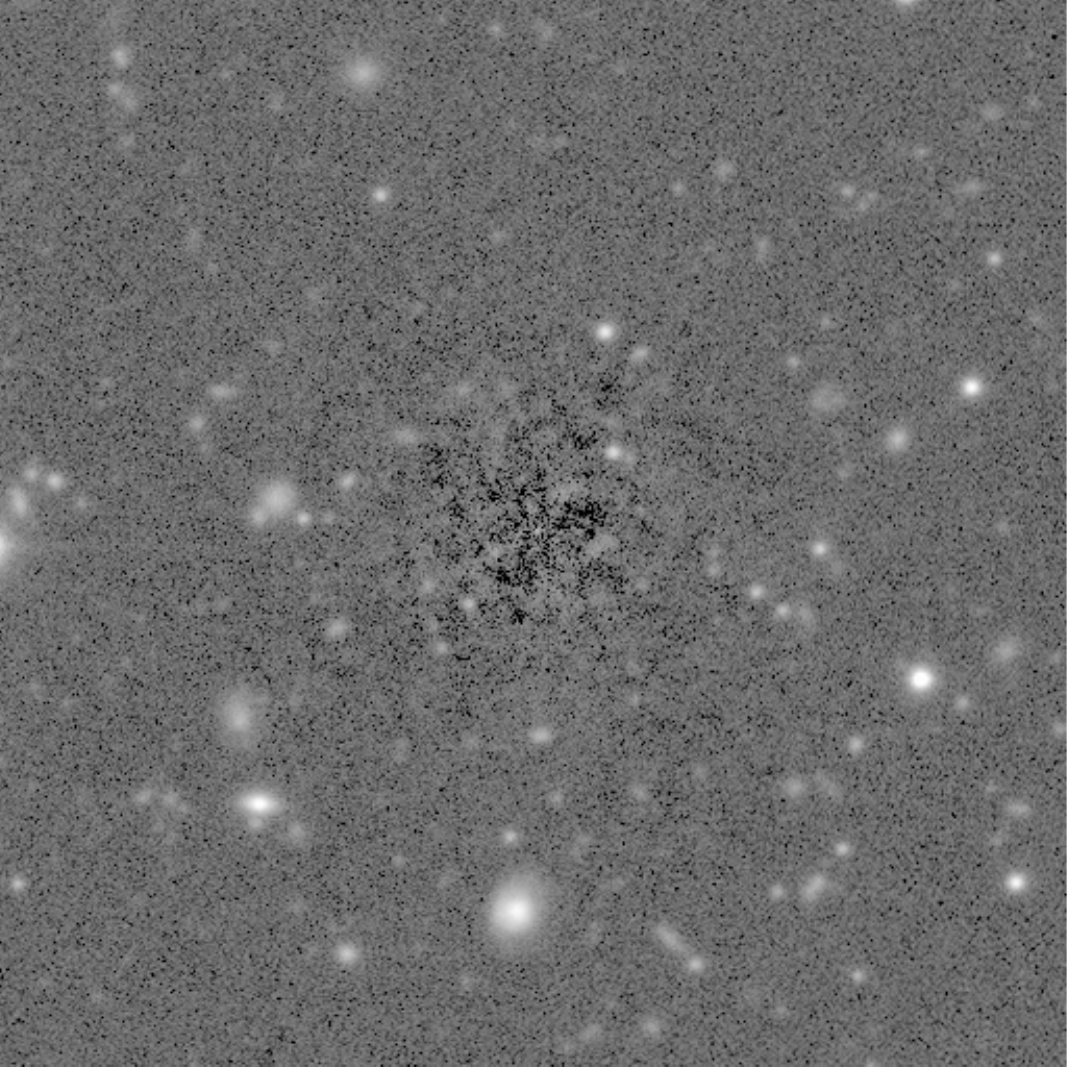}}}
\hfill
\subfigure{\parbox[b]{0.30\textwidth}{\centering
    \text{DAE Residual}\\
    \includegraphics[width=0.29\textwidth]{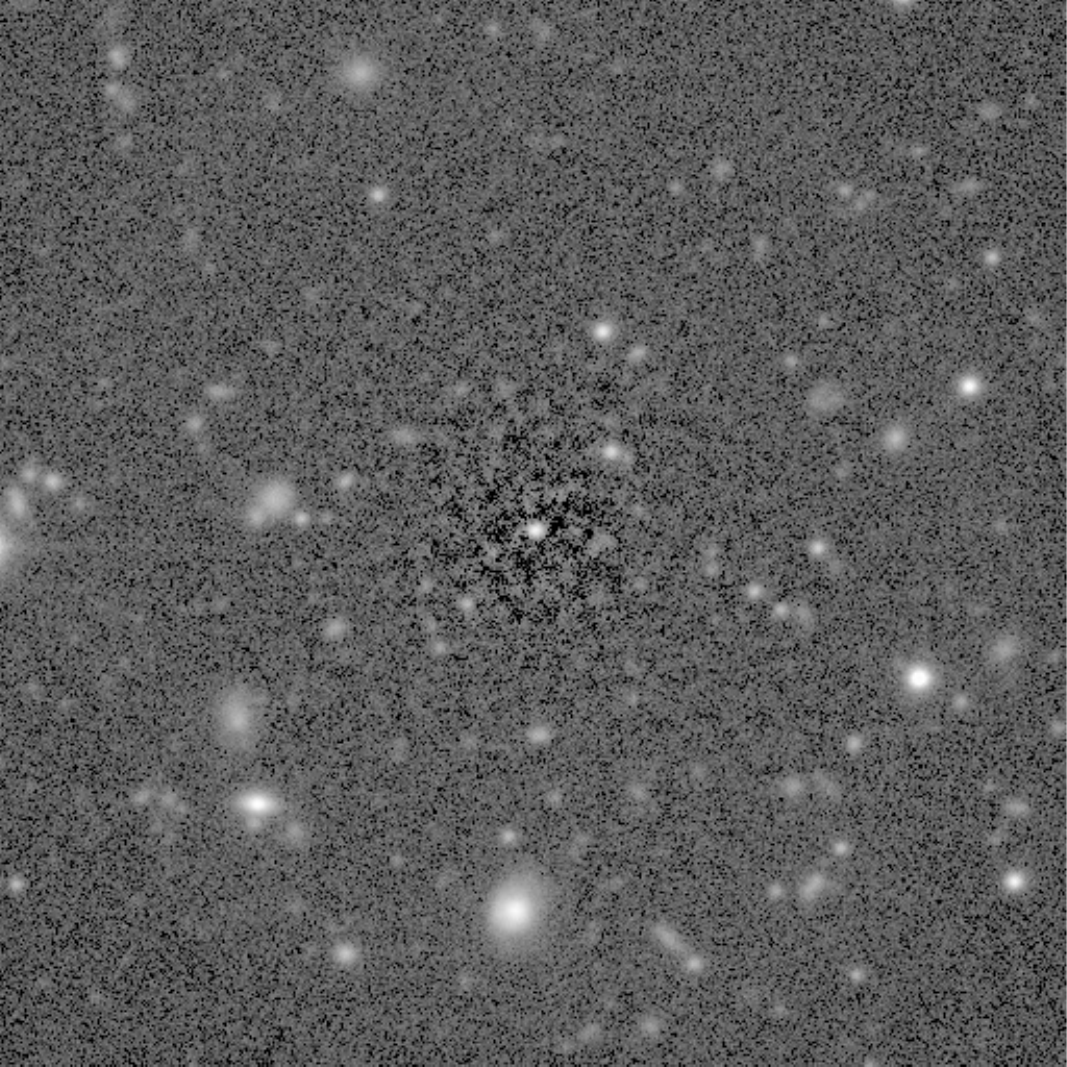}}}

\subfigure{\parbox[b]{0.03\textwidth}{\raggedleft(b)}}
\label{fig:5b}
\hfill
\subfigure{\parbox[b]{0.30\textwidth}{\centering
    \includegraphics[width=0.29\textwidth]{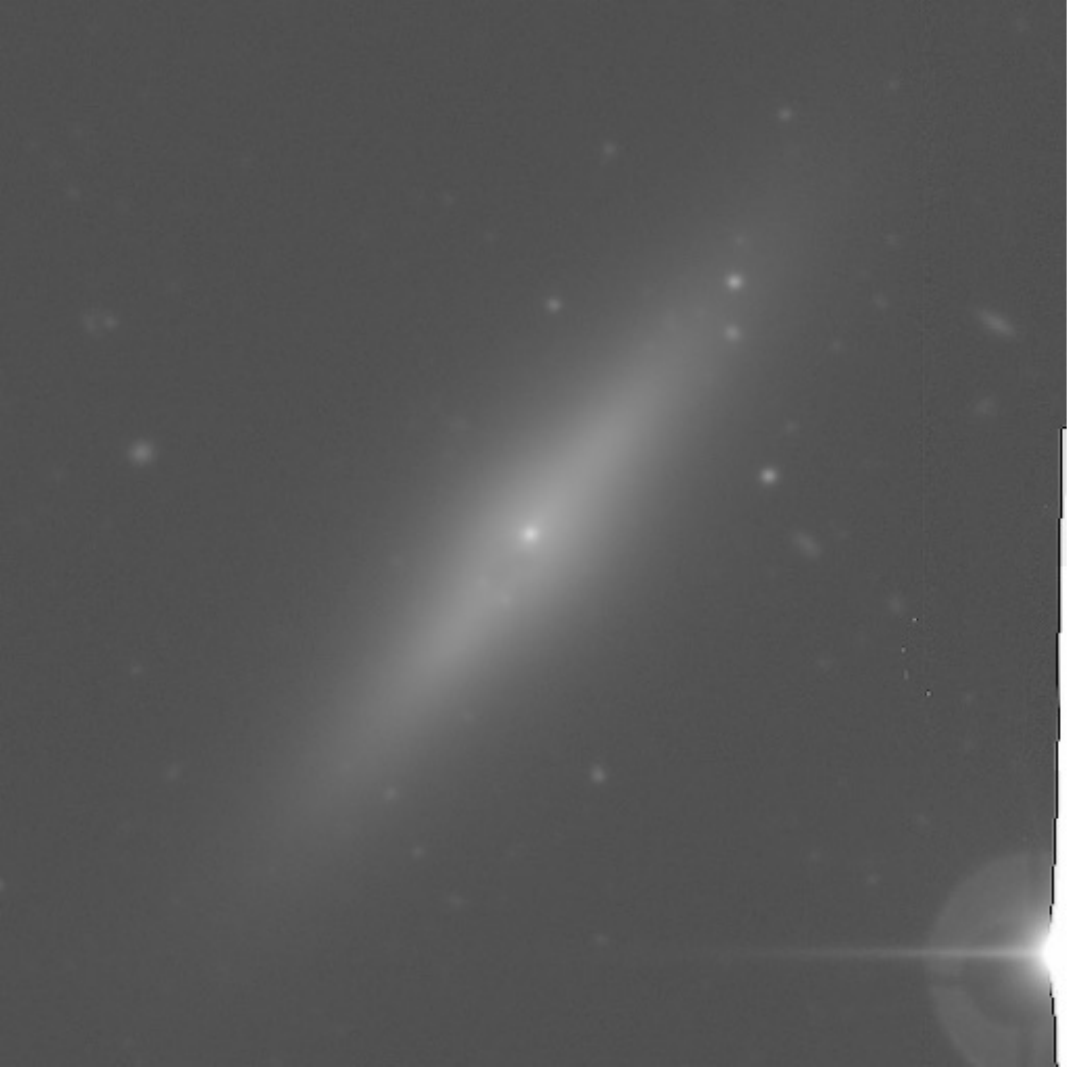}}}
\hfill
\subfigure{\parbox[b]{0.30\textwidth}{\centering
    \includegraphics[width=0.29\textwidth]{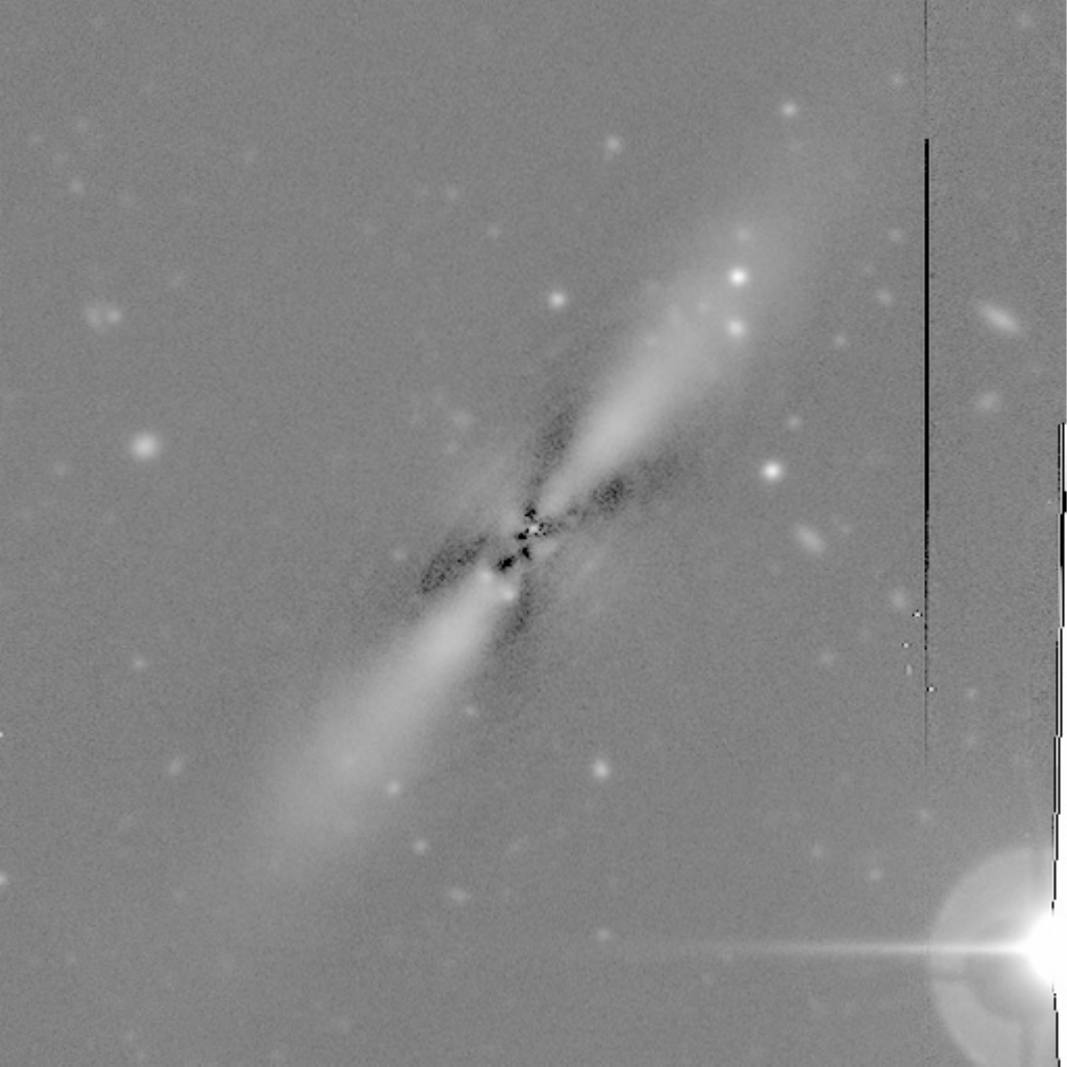}}}
\hfill
\subfigure{\parbox[b]{0.30\textwidth}{\centering
    \includegraphics[width=0.29\textwidth]{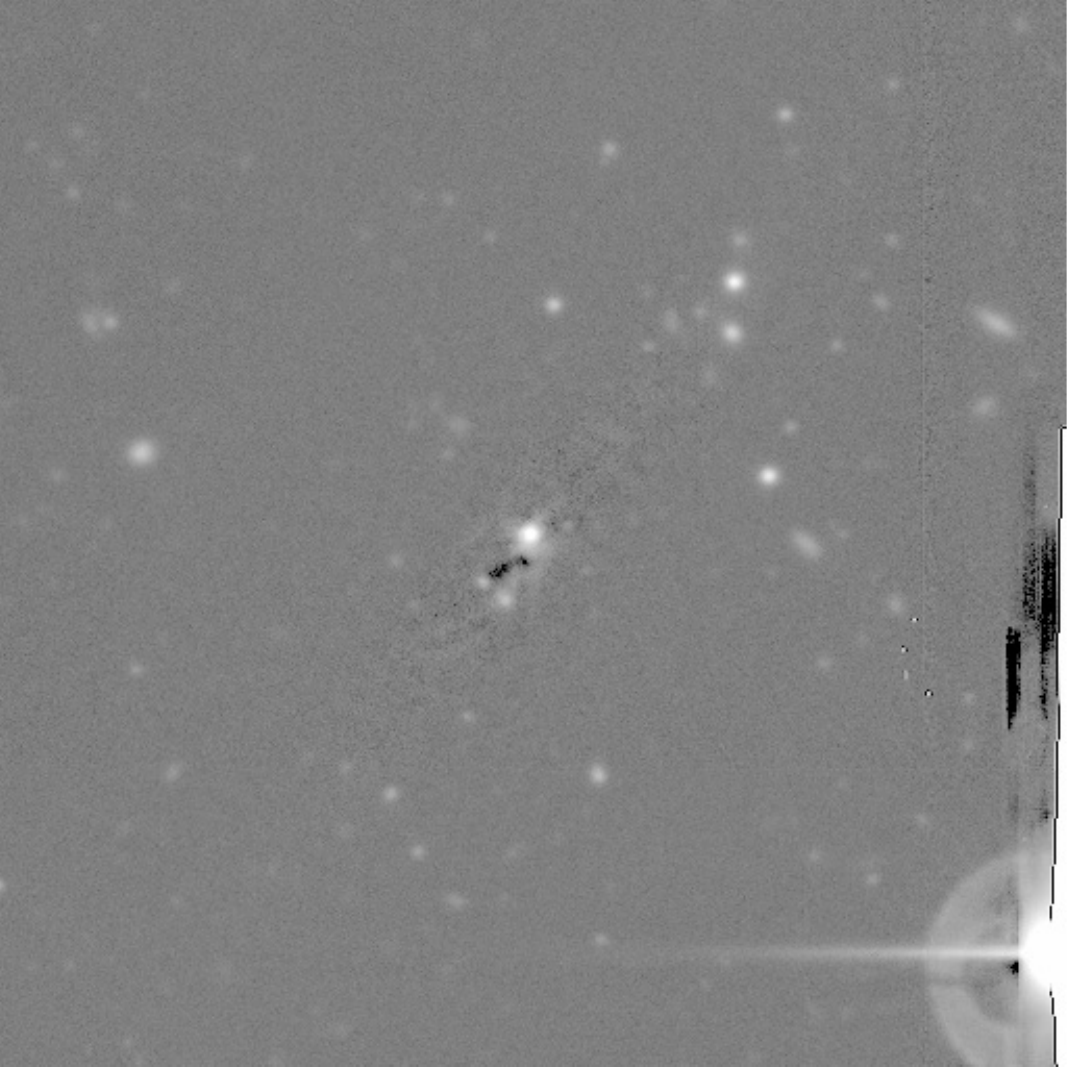}}}

\subfigure{\parbox[b]{0.03\textwidth}{\raggedleft(c)}}
\label{fig:5c}
\hfill
\subfigure{\parbox[b]{0.30\textwidth}{\centering
    \includegraphics[width=0.29\textwidth]{original_12.20.18_.pdf}}}
\hfill
\subfigure{\parbox[b]{0.30\textwidth}{\centering
    \includegraphics[width=0.29\textwidth]{ellipse_12.20.18_.pdf}}}
\hfill
\subfigure{\parbox[b]{0.30\textwidth}{\centering
    \includegraphics[width=0.29\textwidth]{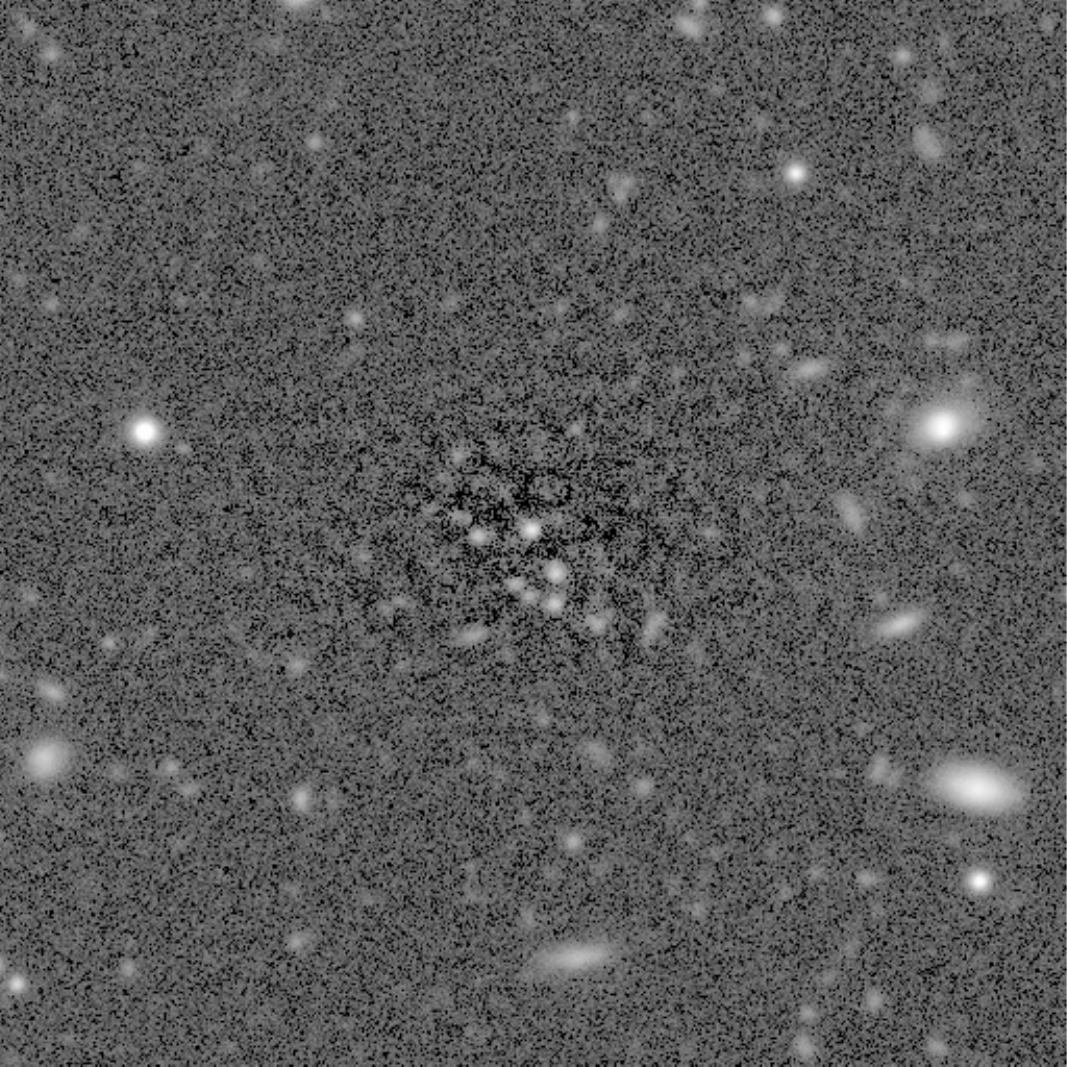}}}

\subfigure{\parbox[b]{0.03\textwidth}{\raggedleft(d)}}
\label{fig:5d}
\hfill
\subfigure{\parbox[b]{0.30\textwidth}{\centering
    \includegraphics[width=0.29\textwidth]{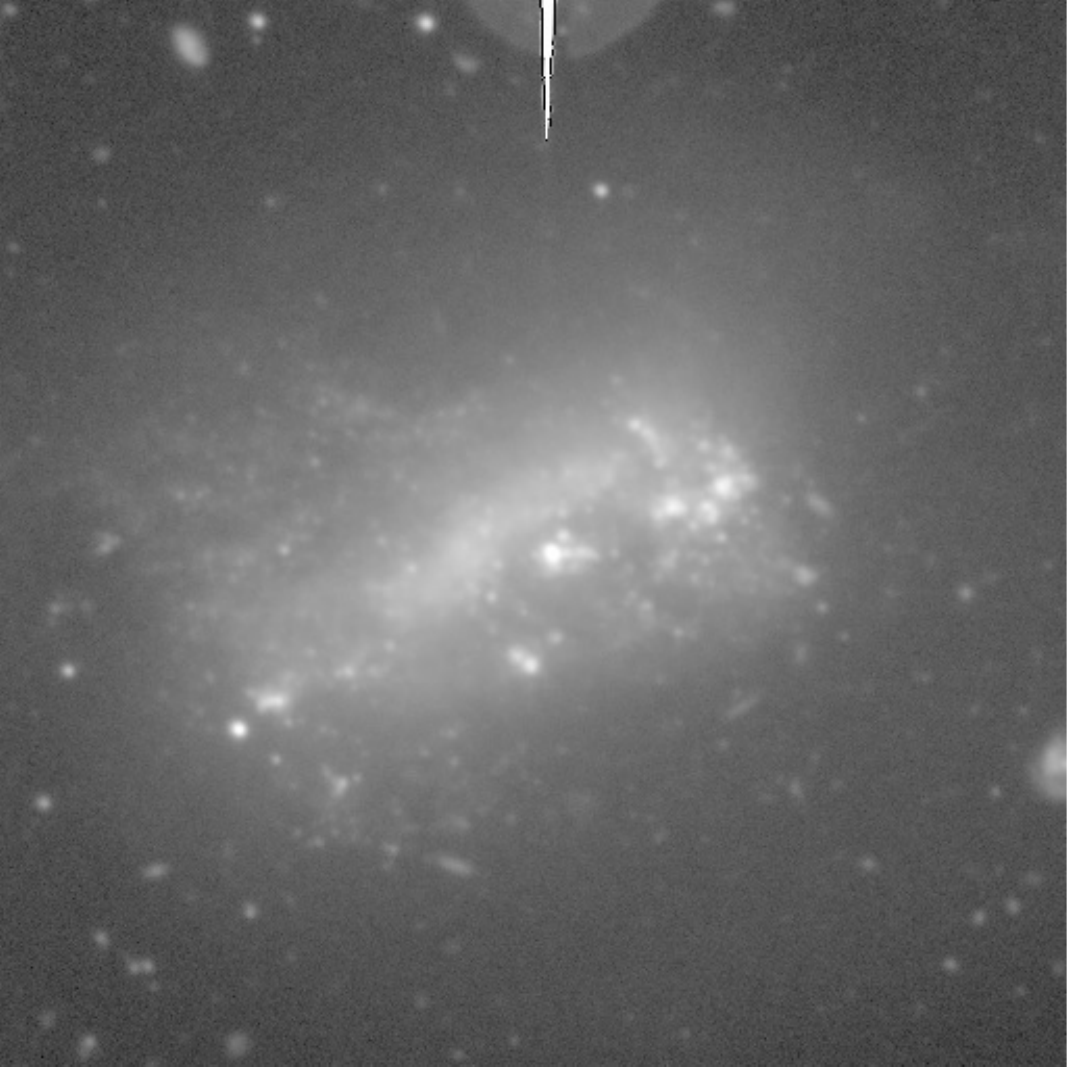}}}
\hfill
\subfigure{\parbox[b]{0.30\textwidth}{\centering
    \includegraphics[width=0.29\textwidth]{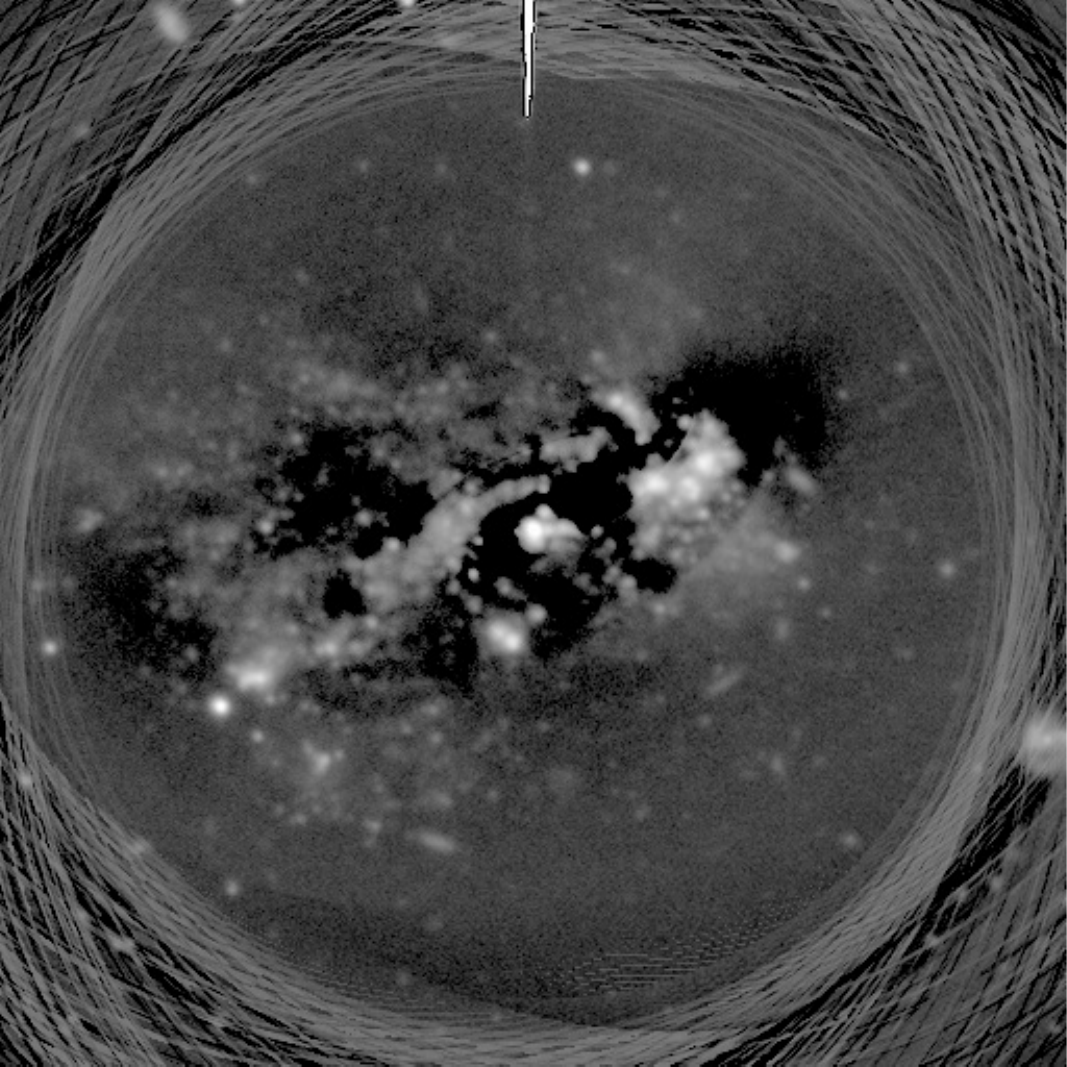}}}
\hfill
\subfigure{\parbox[b]{0.30\textwidth}{\centering
    \includegraphics[width=0.29\textwidth]{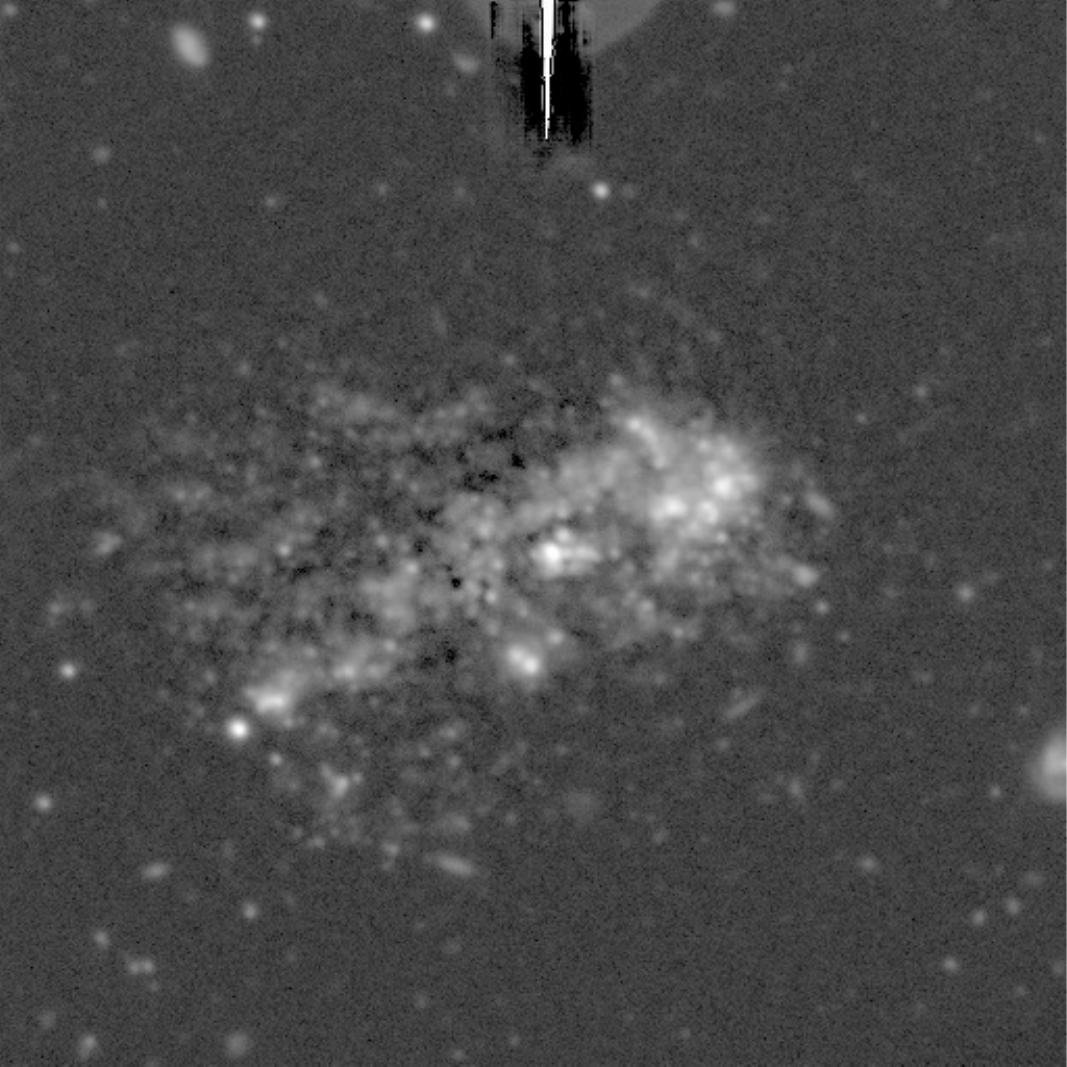}}}

\caption{Test result comparisons on Virgo Cluster galaxies in apparent magnitude range $14{-}15$. 
(a) VCC 0033, a smooth dwarf elliptical galaxy ($M_{g'} = -16.23$). 
(b) VCC 1304, an edge-on galaxy ($M_{g'} = -16.10$). 
(c) VCC 0407, a spiral galaxy with $M_{g'} = -16.83$. 
(d) VCC 1725, a star-forming galaxy ($M_{g'} = -16.80$) with significant artifacts with ellipse model subtraction. 
Ellipse subtraction and the DAE perform similarly well on smooth galaxies, but the DAE yields cleaner residuals for edge-on, spiral, and star-forming galaxies.}
\label{fig:all_examples_14_15}
\end{figure*}
\endgroup
\begingroup
\setlength{\textfloatsep}{0pt} 

\begin{figure*}
\centering

\subfigure{\parbox[b]{0.03\textwidth}{\raggedleft(a)}}
\label{fig:6a}
\hfill
\subfigure{%
    \parbox[b]{0.29\textwidth}{%
        \centering
        \text{Original Galaxy Image}\\
        \includegraphics[width=0.29\textwidth]{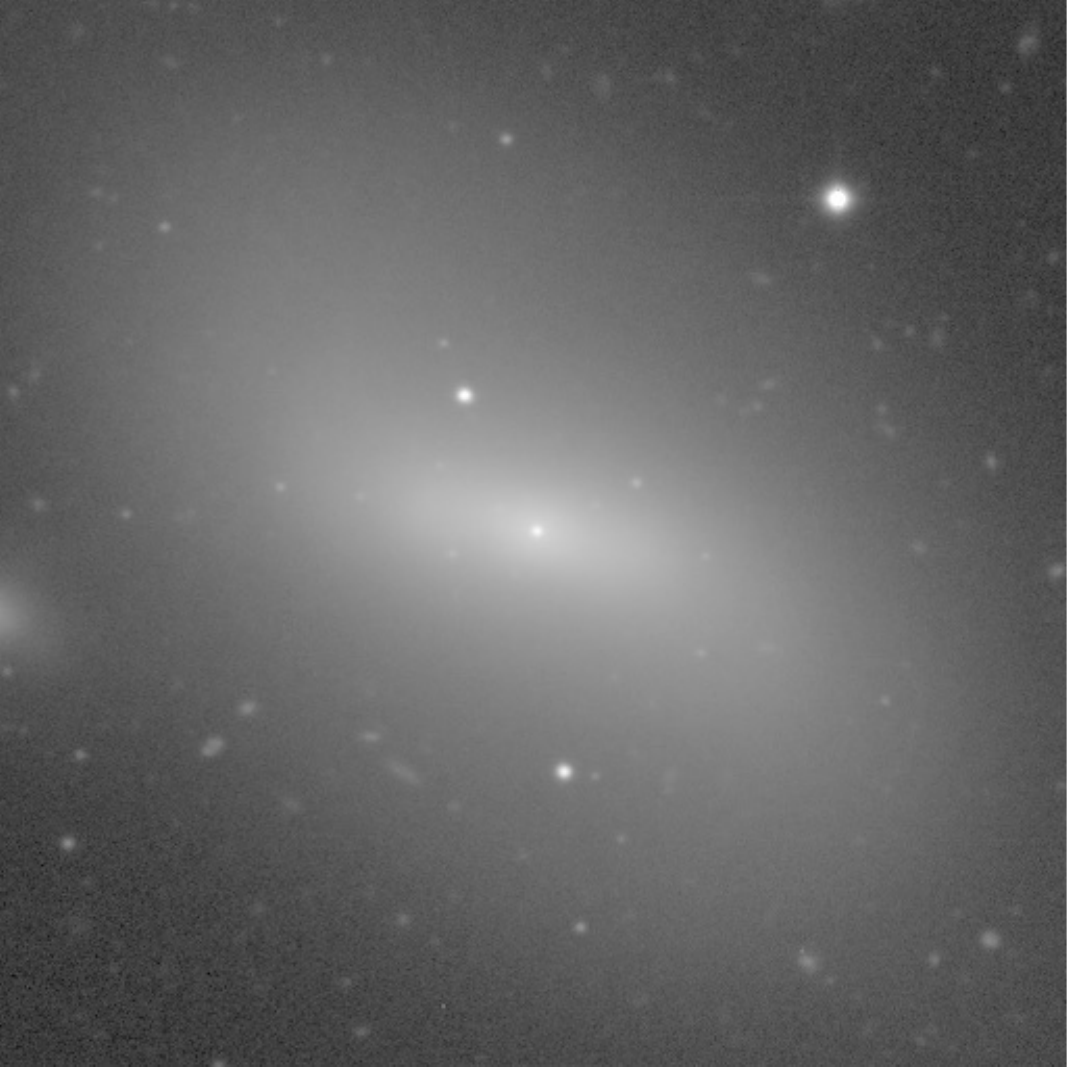}
    }
}
\hfill
\subfigure{%
    \parbox[b]{0.29\textwidth}{%
        \centering
        \text{Ellipse Subtraction Residual}\\
        \includegraphics[width=0.29\textwidth]{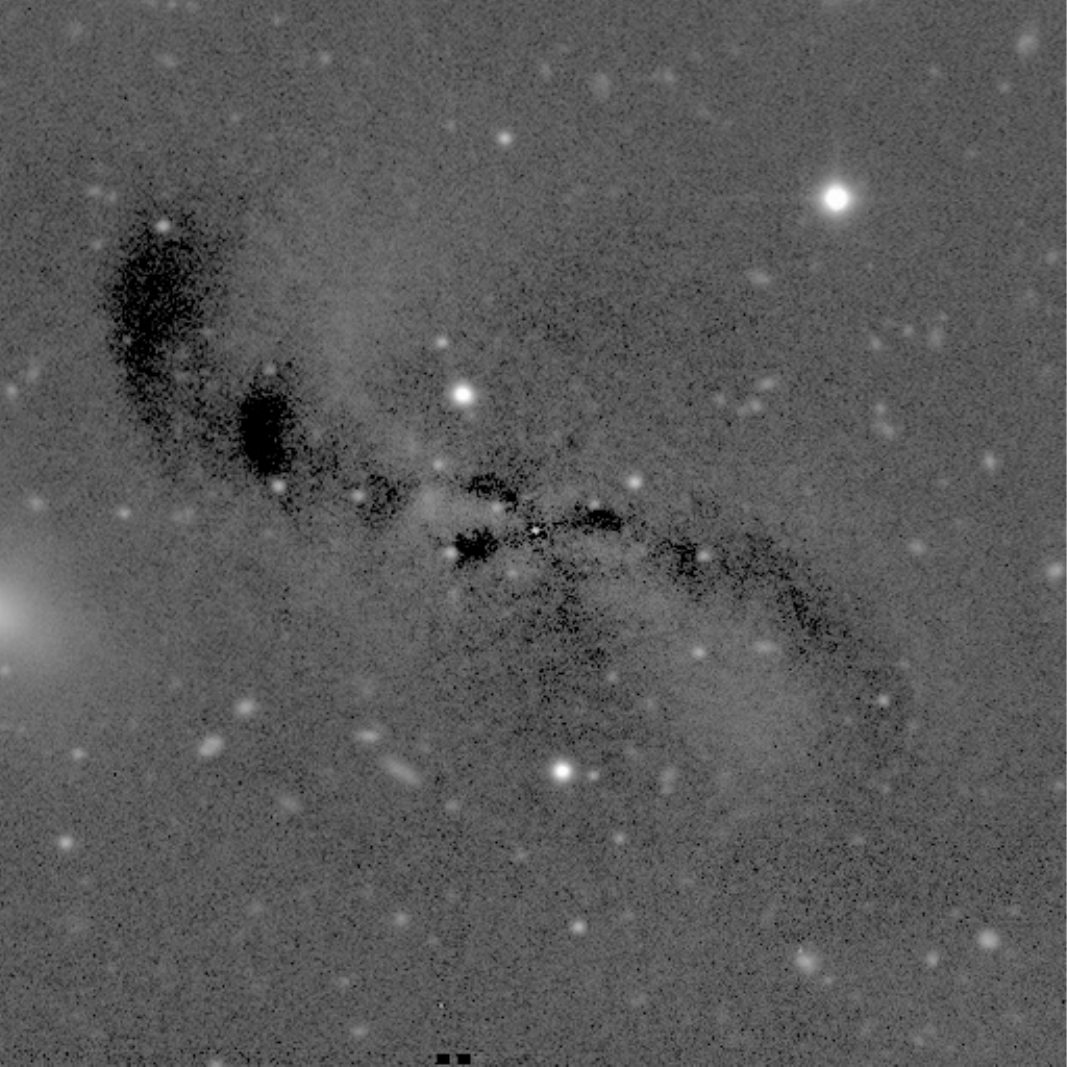}
    }
}
\hfill
\subfigure{%
    \parbox[b]{0.29\textwidth}{%
        \centering
        \text{DAE Residual}\\
        \includegraphics[width=0.29\textwidth]{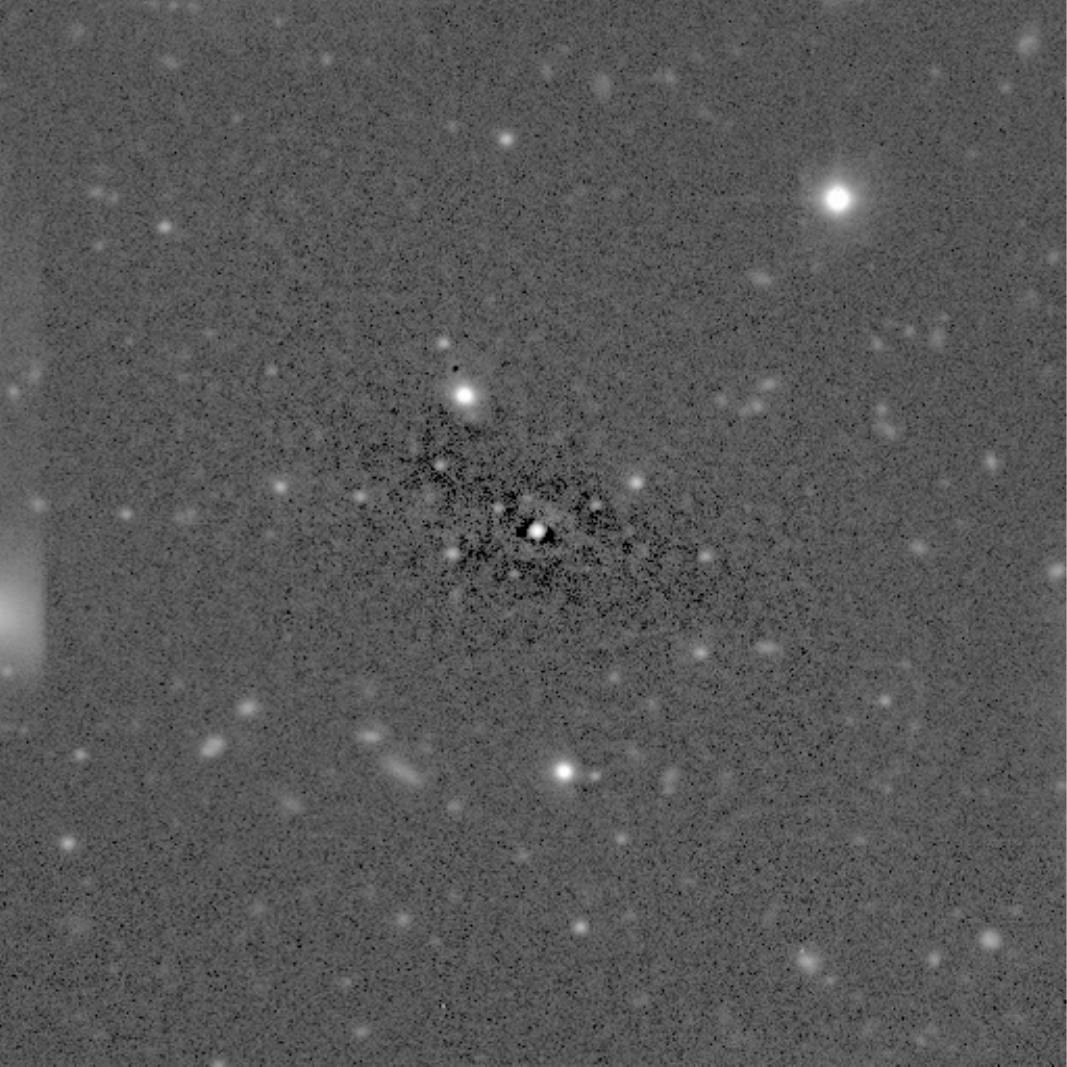}
    }
}

\subfigure{\parbox[b]{0.03\textwidth}{\raggedleft\textbf{(b)}}}
\label{fig:6b}
\hfill
\subfigure{%
    \parbox[b]{0.29\textwidth}{%
        \centering
        \includegraphics[width=0.29\textwidth]{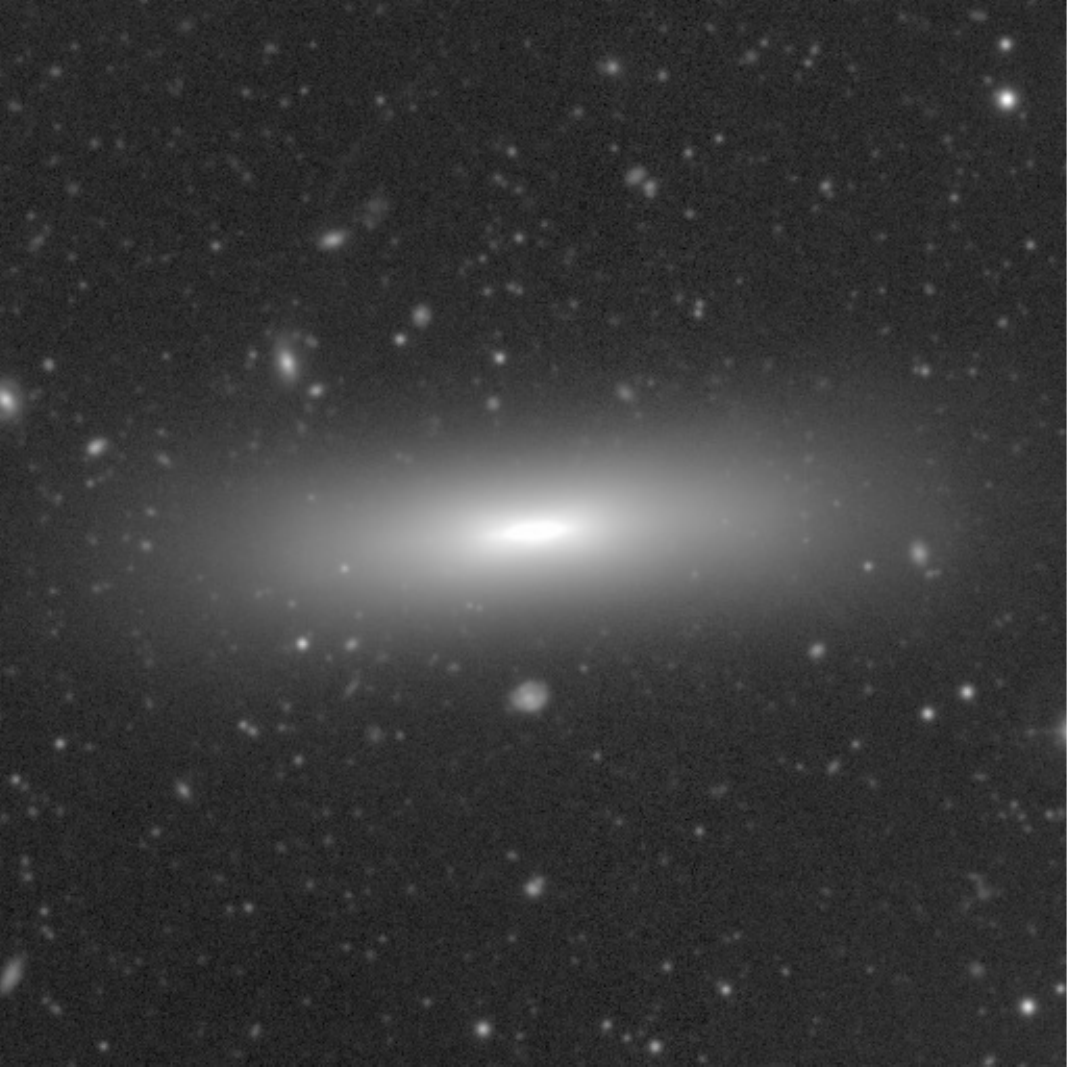}
    }
}
\hfill
\subfigure{%
    \parbox[b]{0.29\textwidth}{%
        \centering
        \includegraphics[width=0.29\textwidth]{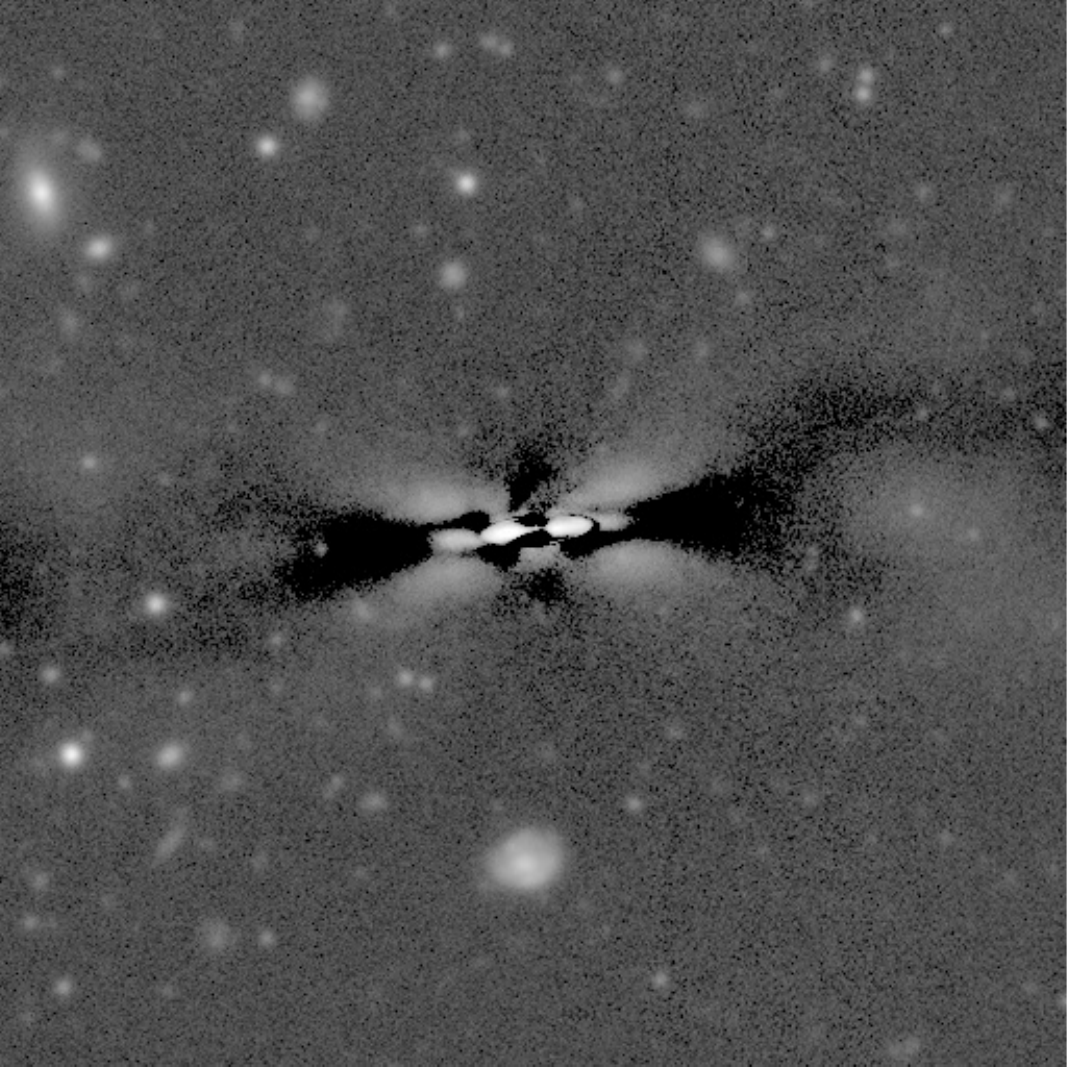}
    }
}
\hfill
\subfigure{%
    \parbox[b]{0.29\textwidth}{%
        \centering
        \includegraphics[width=0.29\textwidth]{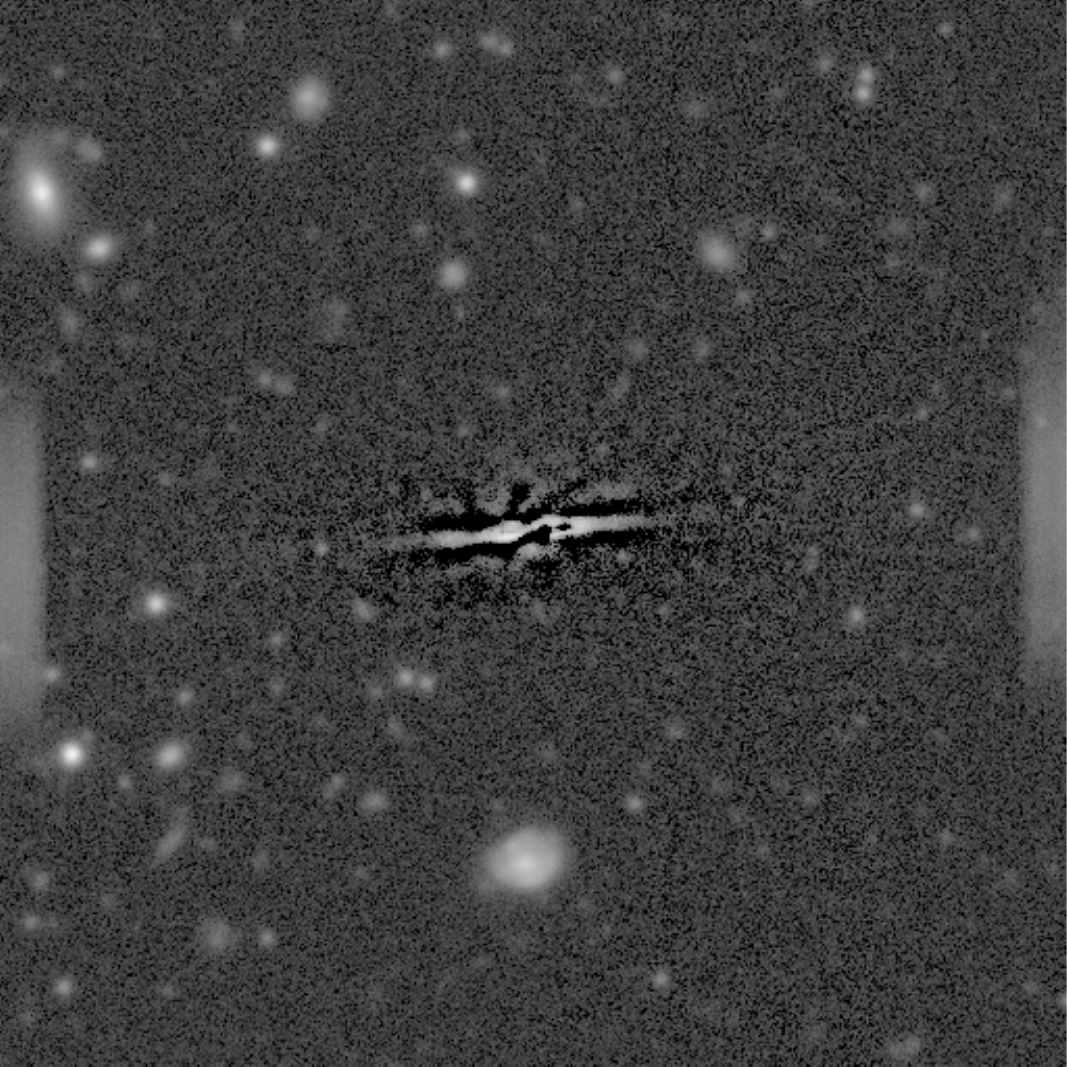}
    }
}

\subfigure{\parbox[b]{0.03\textwidth}{\raggedleft\textbf{(c)}}}
\label{fig:6c}
\hfill
\subfigure{%
    \parbox[b]{0.29\textwidth}{%
        \centering
        \includegraphics[width=0.29\textwidth]{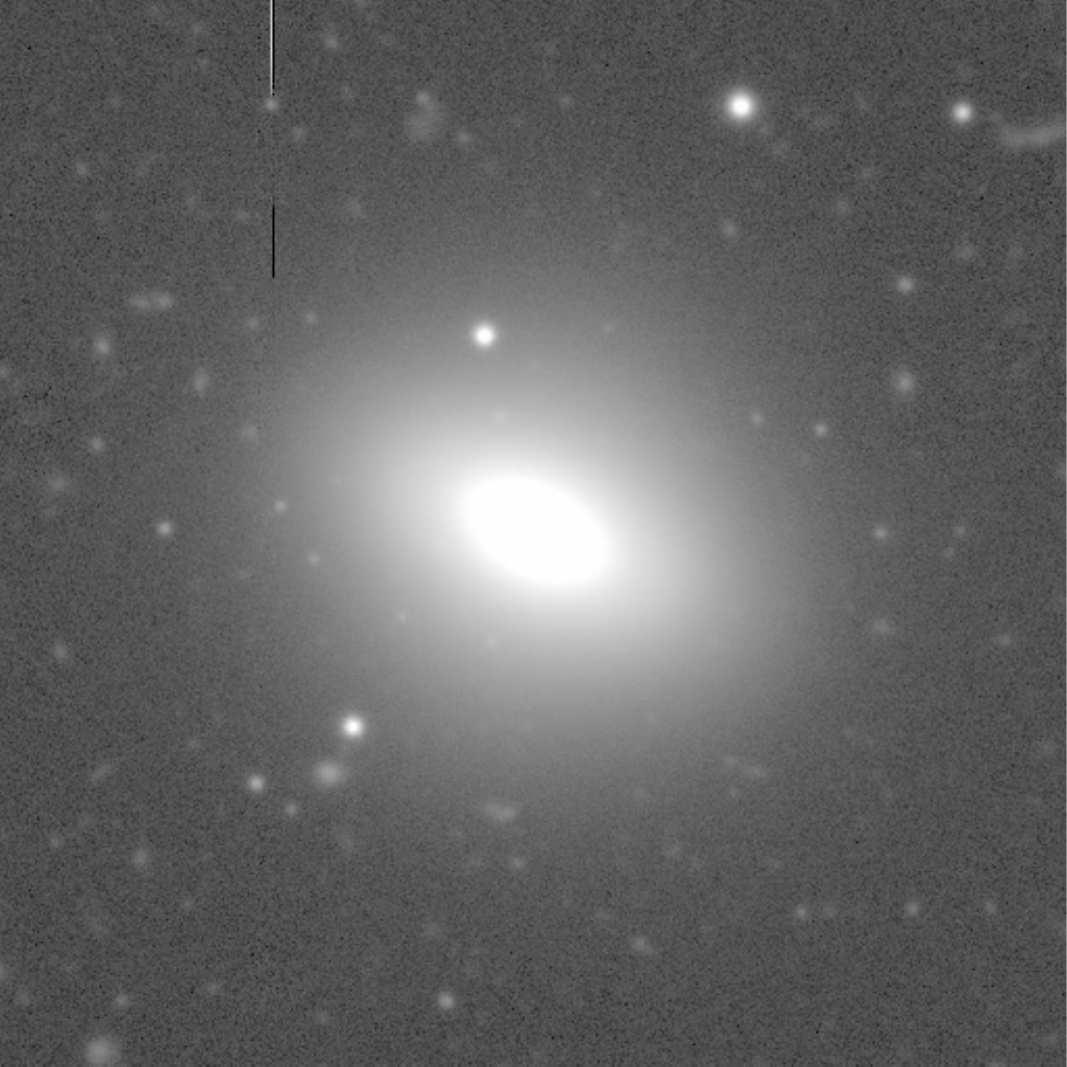}
    }
}
\hfill
\subfigure{%
    \parbox[b]{0.29\textwidth}{%
        \centering
        \includegraphics[width=0.29\textwidth]{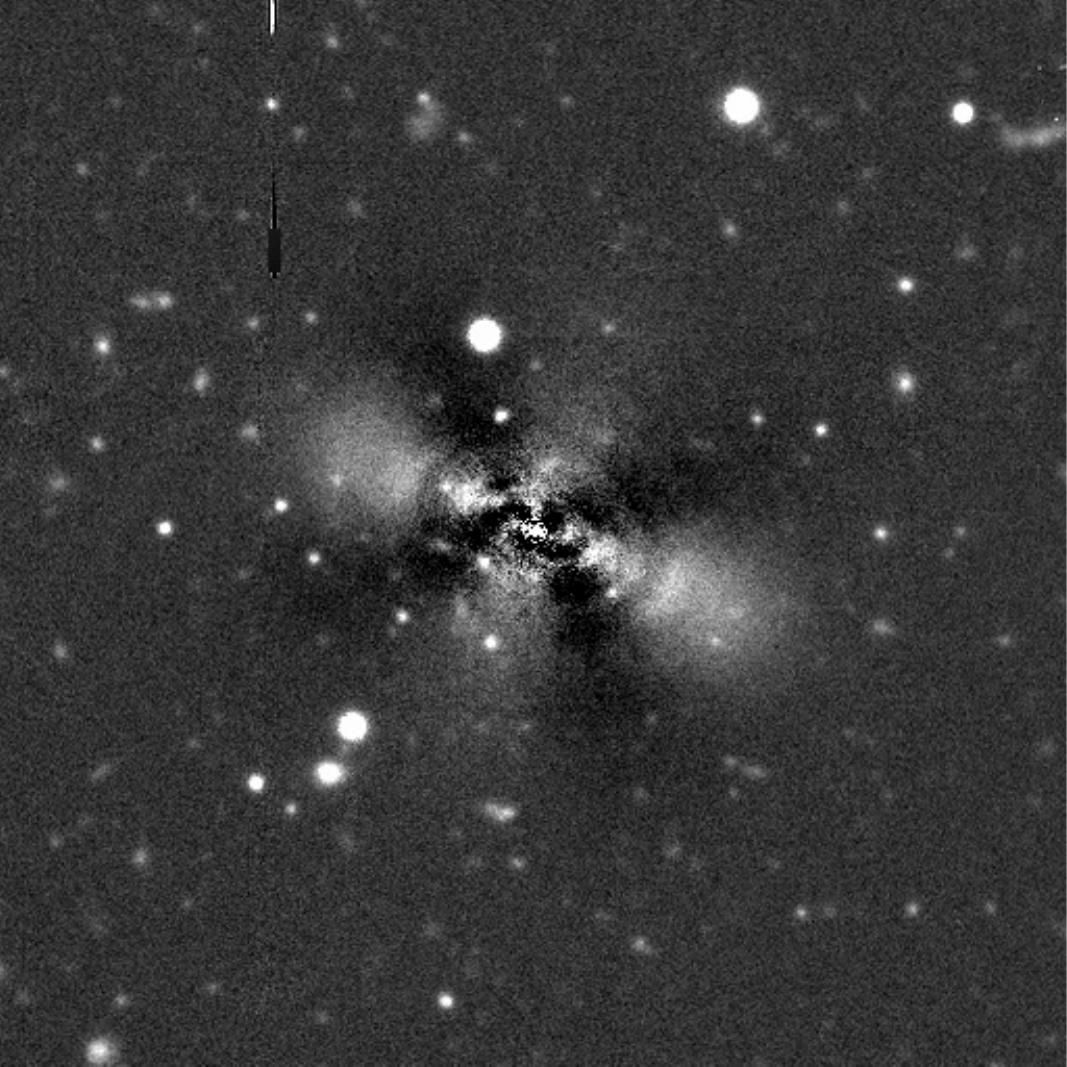}
    }
}
\hfill
\subfigure{%
    \parbox[b]{0.29\textwidth}{%
        \centering
        \includegraphics[width=0.29\textwidth]{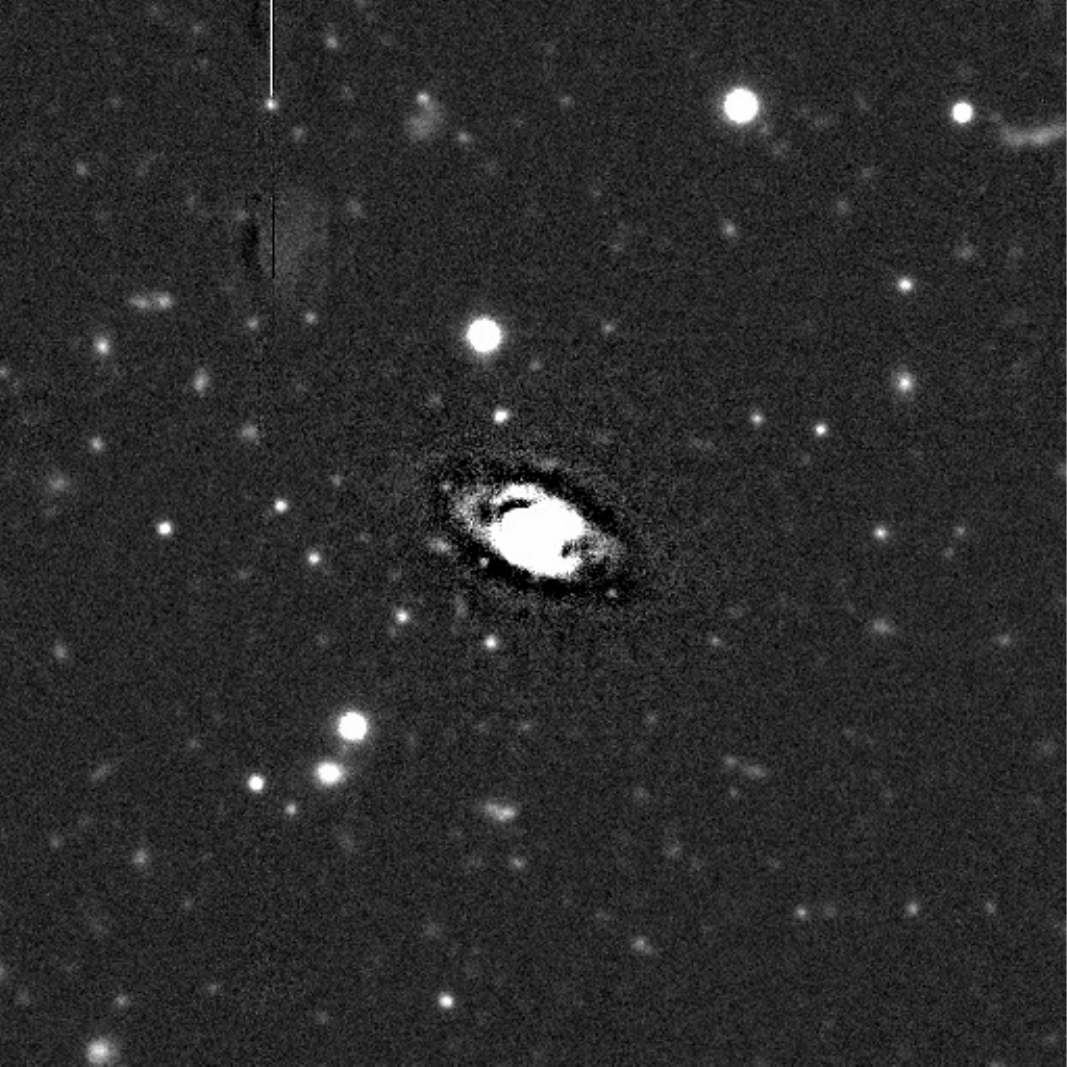}
    }
}

\caption{Test result comparisons on Virgo Cluster galaxies in apparent magnitude range $13<g'<14$~mag. 
(a) VCC 1949, a galaxy with weak spiral structure  ($M_{g'} = -17.43$). 
(b) VCC 0672, an edge-on galaxy ($M_{g'} = -17.48$). 
(c) VCC 0575, a galaxy with a bulge ($M_{g'} = -17.52$). The denoising autoencoder reduces the X-shaped residuals, but struggles to fit the central bulge, leaving a bright central region in the residual image.}
\label{fig:vcc_13to14_examples}
\end{figure*}

\endgroup

\subsection{Testing for Photometric Bias}
Clean subtraction of a smooth background or galaxy is only useful if it does not compromise the photometry of the revealed sources. Local, non-parametric background subtraction, such as ring median or other spatial filters, can often bias photometry. This is because filtering on smaller scales to better follow the underlying light leads to source contamination in background estimates, which in turn leads to overestimation of the local background and underestimation of source flux. Isophote fitting techniques get around this problem by taking advantage of the elliptical symmetry of galaxies to use infomation from parts of the image that are well away from sources to estimate the local background. This only works, however, if the assumption of elliptical isophotes is valid. The denoising autoencoder still uses the finite parameter space spanned by galaxies in their shapes and surface brightness profiles to estimate the spatially varying light profile, allowing for many more degrees of freedom, but is computationally fast. Here, we perform a test to see if photometry on residual images produced by the denoising autoencoder are in any way biased.

\begin{figure}[h]
    \subfigure[The difference in integrated magnitudes extracted from residual images produced by autoencoder and ellipse fitting vs aperture diameters.]{%
        \includegraphics[width=0.95\linewidth]{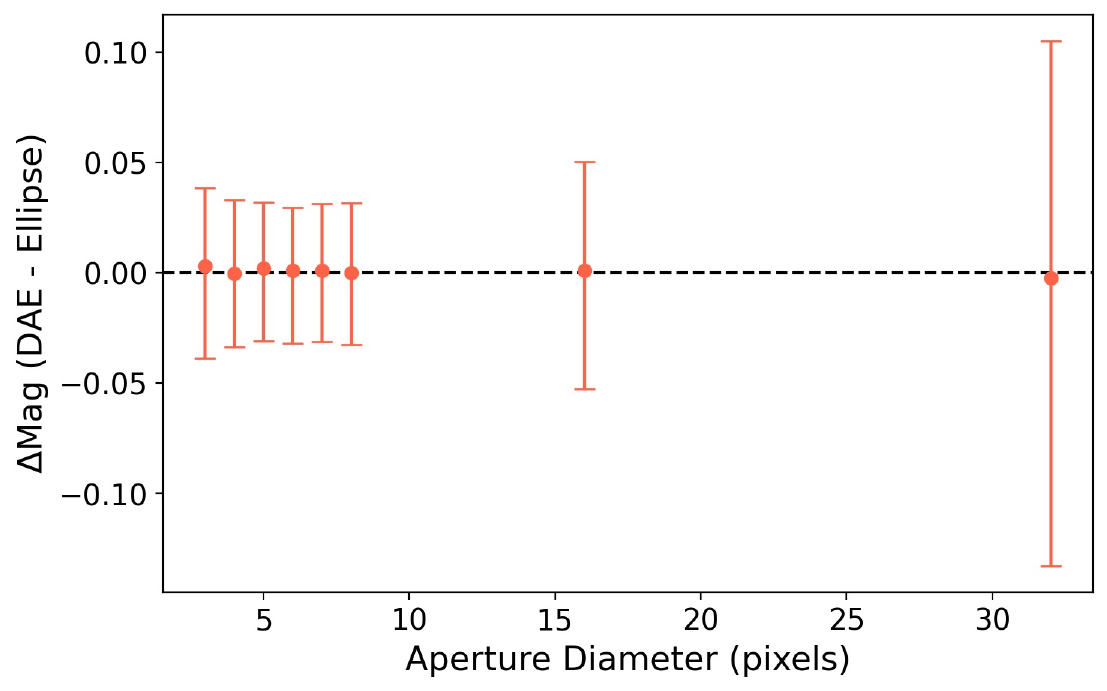}
        \label{fig:enc_ell_mag}
    }
    \vfill
    \subfigure[The difference in integrated magnitudes extracted from residual images produced by ring median filter (inner radius = 2 pixels / 0.374 arcsecs, outer radius = 3 pixels / 0.561 arcsecs) and ellipse fitting vs aperture diameters.]{%
        \includegraphics[width=0.95\linewidth]{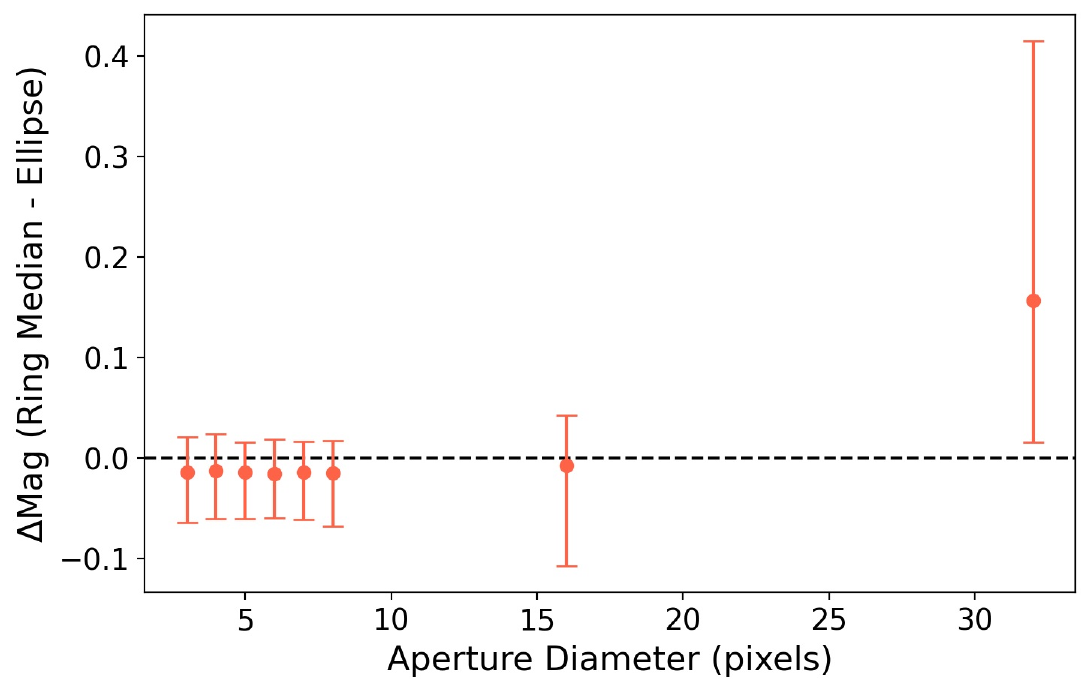}
        \label{fig:ring_ell_mag}
    }
    \caption{For each galaxy model subtraction method used, we extracted the integrated magnitude of objects in the residual image across different aperture sizes and present the comparison. The top image shows 
    The lower and upper error bars represent the 16th and 84th percentiles of the data points. The magnitude of objects obtained using different aperture sizes can vary. This plot shows that the magnitudes obtained by the denoising autoencoder are consistent with those obtained from ellipse fitting and are not dependent on aperture size. In contrast, the ring median filter subtracted images show over-subtraction for small aperture sizes and under-subtraction for large aperture sizes.}
    \label{fig:autoencoder_ellipse}
\end{figure}

To evaluate whether our model performs well on this front, we run Source Extractor on the residual images of the denoising autoencoder and those of both ellipse fitting and ring median filtering. We then match the \textbf{detected sources} in each of the residual images by their positions and compare their magnitudes observed with apertures of different sizes. The medians of $\Delta$Mag~(autoencoder$-$ellipse) at each aperture size are shown in Figure \ref{fig:autoencoder_ellipse}, with the lower and upper error bars representing 16th and 84th percentiles of the data points. Larger apertures are more sensitive to incorrect background estimations, so biases that correlate with aperture size are often signs of a problem with background estimation. 
Figure \ref{fig:enc_ell_mag} shows that the medians of $\Delta$Mag all lie close to the 0-valued horizontal line, indicating that the galaxy model subtraction results produced by the autoencoder is consistent with the subtraction results from ellipse fitting. Also, our autoencoder model maintains a consistent performance across a range of aperture sizes from 3 pixels to 32 pixels. 
We also run Source Extractor on the residuals produced by ring median filter and made the same comparison to the ellipse fitting results. Figure \ref{fig:ring_ell_mag} clearly shows the limitation of ring median filter; it consistently over-subtracts for small apertures and under-subtracts for the largest one.

\bfseries
\subsection{Injection-Recovery Test}
\subsubsection{Test Setup}
To further evaluate the quality of the denoising autoencoder subtraction compared to ellipse subtraction, we perform an injection-recovery test with mock globular clusters (GCs). We simulate GCs with Point Spread Functions (PSFs) at the location of each galaxy and add them into the residual images produced by ellipse fitting and denoising autoencoder. If there are left-over galaxy structures in the residual image, the recovery rates of these mock globular clusters will be affected. 

The radial distances of GCs relative to their host galaxy centers are sampled from a probability distribution  $p(r) \propto I(r)\, r$, where $I(r)$ is the Sérsic surface brightness profile

\begin{equation}
I(r) = I_0 \exp\left[-b_n\left(\left(\frac{r}{R_{e,\mathrm{gc}}}\right)^{1/n} - 1\right)\right]
\label{eq:sersic}
\end{equation}

and where we choose Sérsic index $n =2$ with $b_n\approx3.672$ \citep{1999A&A...352..447C}. The azimuthal angle $\phi$ of the GCs are drawn random uniformly from the range $[0,2\pi]$ so the angular distribution is symmetric. 

The effective radius of each GC system is taken from the broken power-law function for the scaling relation between GC system size and galaxy stellar mass presented in \citet{Lim_2024}:
\begin{equation}
    R_{e, \mathrm{GC}}= R_p\left(\frac{M_*}{M_p}\right)\left[\frac{1}{2}\left\{1+\left(\frac{M_*}{M_p}\right)\right\}\right]
\end{equation}
where $M_p$ is the pivot mass at which the function slope changes, $R_p$ is the radius  at $M_p$,  $\alpha$ is the slope at the low-mass end, $\beta$ is the slope at the high-mass end, and $\delta$ is a smoothing factor. We use smoothing factor $\delta = 6$ and other parameter values as  fitted in \citet{Lim_2024} ( ``All-GCs"): $M_p = 6.5\times 10^{10}M_\odot$, $R_p = 8.3\text{ kpc}$, $\alpha = 0.34$, $\beta = 1.30$. We convert $R_{e,\mathrm{GC}}$ from physical units (kpc) to pixels using \begin{equation}
    R_{e,\mathrm{GC}}(\mathrm{pix}) = \frac{R_{e,\mathrm{GC}}(\mathrm{kpc}) \times 206265}{D(\mathrm{kpc}) \times s_\mathrm{pix}}
\end{equation}
where 206265 is the number of arcseconds per radian, $D$ is the distance to the Virgo Cluster ($16.5\times 10^3$~kpc) and $s_\mathrm{pix}$ is the pixel scale of our CFHT/MegaCam imaging (0.187 arcsec~pix$^{-1}$): 

We use flux-scaled local PSFs to model realistic GCs in each galaxy. The local PSFs are created using \texttt{DAOPHOT} \citep{Stetson_1987, ngvs} to accurately capture local conditions including both the atmospheric and instrumental profiles. The mock GCs are generated to match observed magnitude and spatial distributions. We sample the apparent magnitude of GCs uniformly in the range $22\le m_{\rm GC}\le26$~mag following the typical brightnesses of GCs observed by NGVS. (We note that GC brightness are typically distributed log-normally, but we chose to sample from a uniform distribution to increase our sampling at the bright and faint ends). For each GC, we use the PSF generated for its host galaxy to represent its shape, and then scale the flux according to its injected magnitude $m_{\rm inj}$ using the standard magnitude-flux conversion law $F = 10^{(30 - m_{\rm inj}) / 2.5}$ where 30 is the zero-point magnitude for NGVS.  

To recover the injected mock GCs, we use Source Extractor  to detect and measure sources in residual images produced by the ellipse fitting and denoising autoencoder, injected with mock GCs. For consistency, we use a fixed 8-pixel-diameter aperture to measure the source magnitudes. Since this aperture cannot capture total flux of the source due to the extended wings of the PSF, we apply an aperture correction on the detections. This correction is done by calculating the magnitude difference between the total integrated PSF flux and the flux within the 8-pixel aperture, then subtracting this difference from detected magnitude. 

We consider a source to be successfully matched to an injected mock GC if the detected source position is within 2 pixels of the injection location and its corrected magnitude is less than 1~mag away from the injected magnitude. We then evaluate the recovery rate and photometric accuracy on results from both subtraction methods. 

We apply this test to the residuals produced by ellipse subtraction and denoising autoencoder on 114 NGVS galaxy images with $14<g'<15$~mag. We inject 20 GCs into each residual image and repeat this test for 1000 iterations on each image. 

\subsubsection{Test Results}






To compare the effectiveness of the two subtraction methods across different galaxy types, we aggregate the recovery results into 5 morphology categories: nucleated elliptical galaxies, non-nucleated elliptical galaxies, smooth spiral galaxies, edge-on galaxies, and star-forming galaxies. The classification is mainly based on the morphological catalog of NGVS galaxies from Kurzner et al. (2025, submitted), with manual classification for edge-on galaxies.  We then visualize the results using a 2D heatmap as a function of injected magnitude and distance to galaxy center (in units of galaxy effective radius, $R_e$). 

\begin{figure*}
    \centering
    \subfigure[Elliptical Galaxies without Nuclei]{%
        \includegraphics[width=1\linewidth]{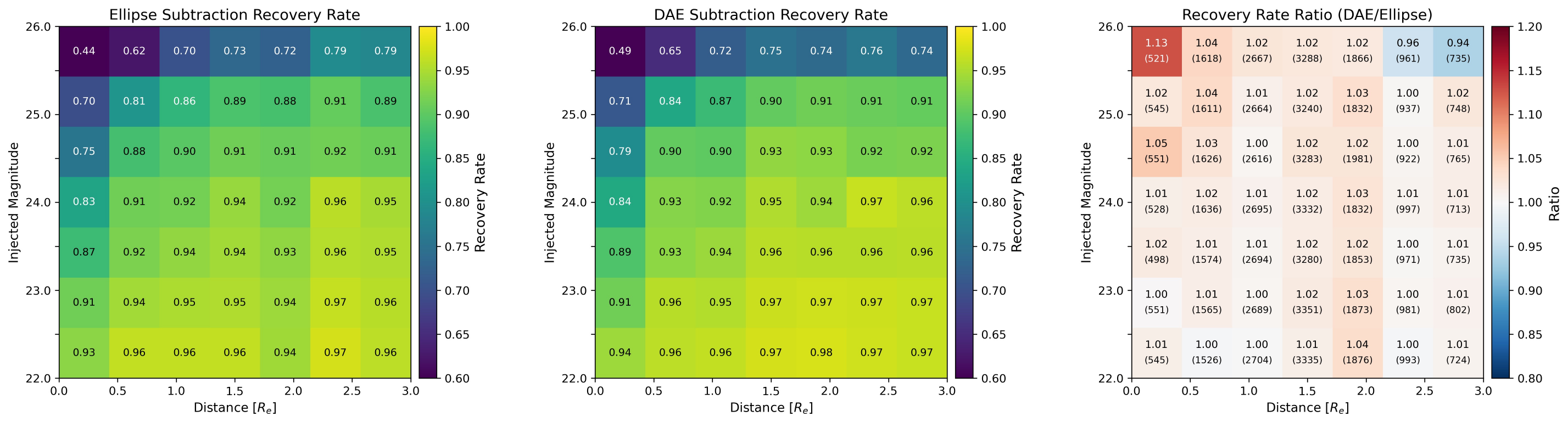}
        \label{fig:heatmap_ellipse_no_nuc}
    }\\[1ex]
    \subfigure[Elliptical galaxies with Nuclei]{%
        \includegraphics[width=\linewidth]{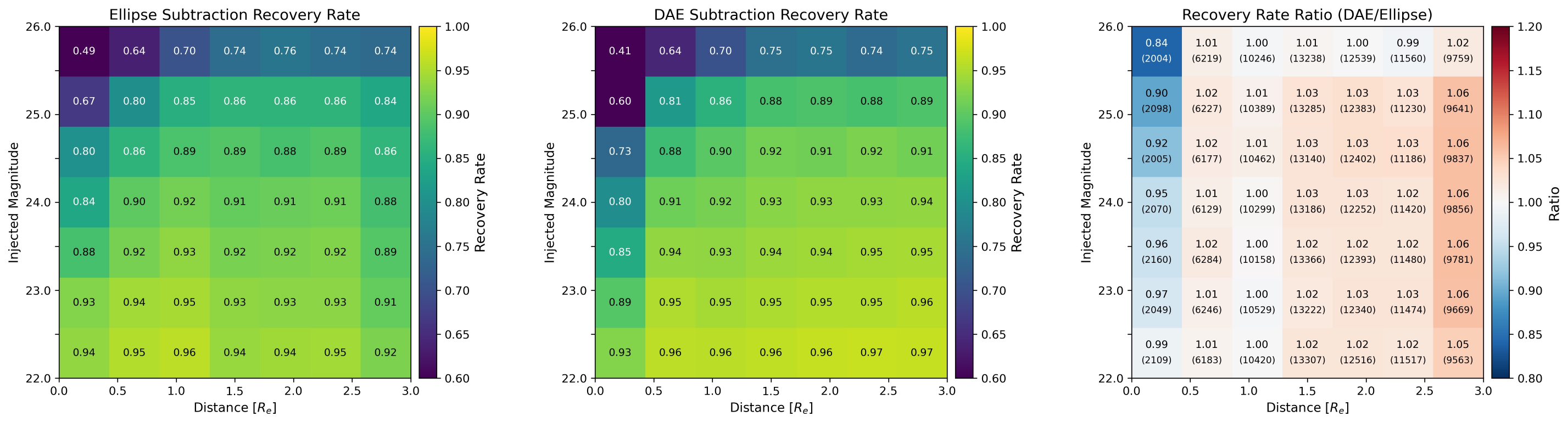}
        \label{fig:heatmap_ellipse_w_nuc}
    }\\[1ex]
    \subfigure[Edge-on Galaxies]{%
        \includegraphics[width=\linewidth]{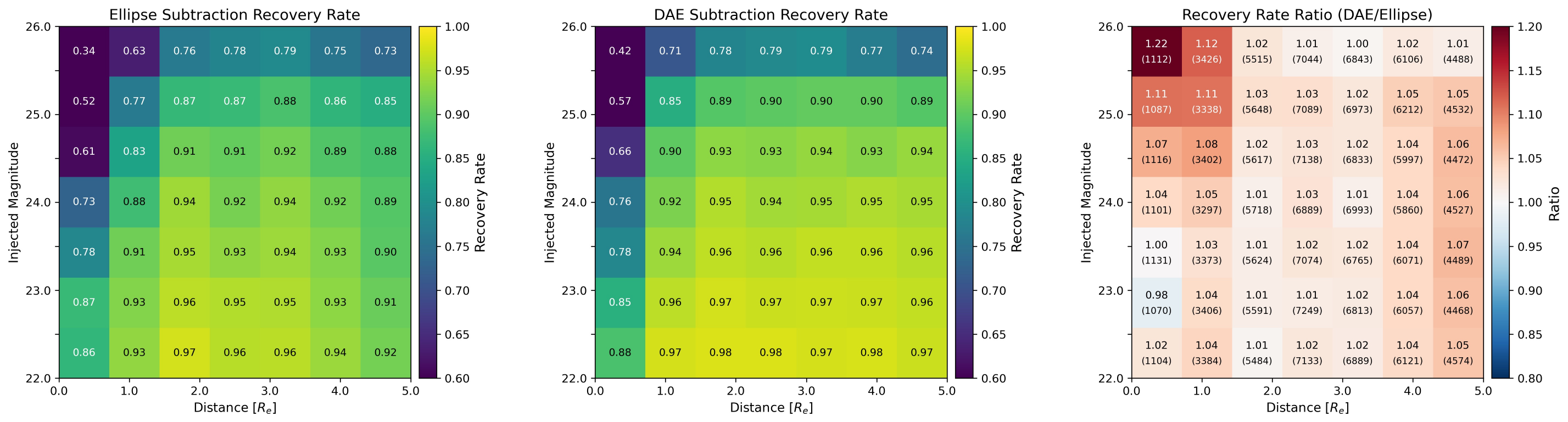}
        \label{fig:heatmap_edge_on}
    }\\[1ex]
    \subfigure[Smooth Spiral Galaxies]{%
        \includegraphics[width=\linewidth]{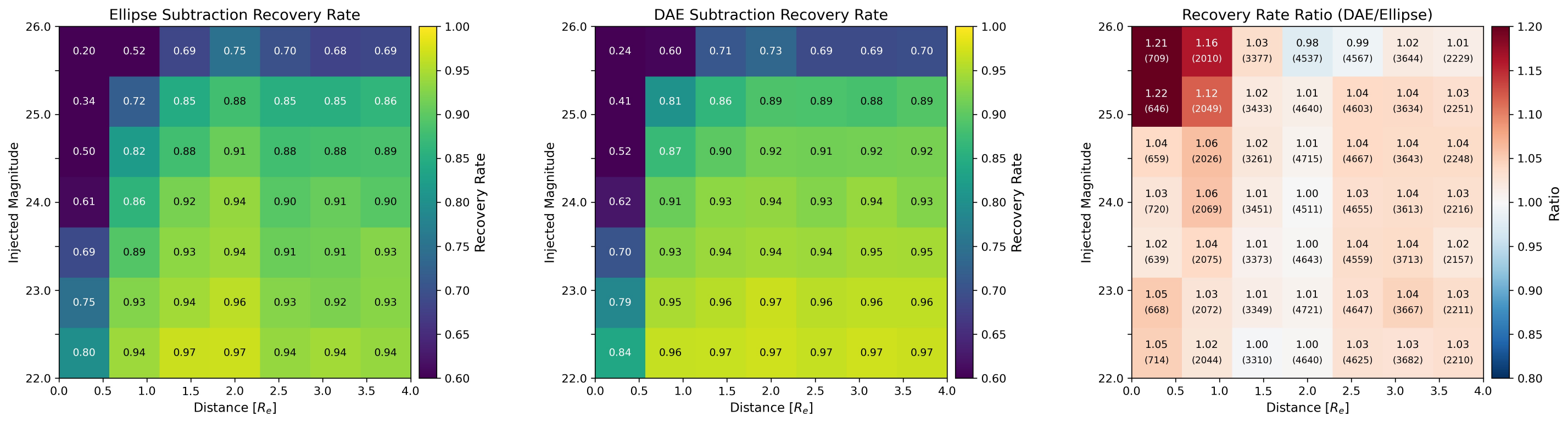}
        \label{fig:heatmap_smooth_spiral}
    }
    \caption{Aggregated recovery rate and mean magnitude error heatmaps created from mock globular clusters injected into residual images for different galaxy morphologies.  The left panel is the recovery rate from ellipse subtraction, the middle panel is recovery rate from DAE, and the right panel shows the ratio between the two (top number in each bin). The bottom number in each bin represents the total number of objects injected in that bin. For all non-nucleated galaxies, particularly those with complex morphology, the recovery rate of faint GCs is markedly better in DAE-subtracted images.}
    \label{fig:heatmap_bundle}
\end{figure*}

When applied to non-nucleated elliptical galaxies, both subtraction methods can remove the smooth galaxy light effectively. The overall recovery rates are consistently high ($>90\%$ for mock GCs with $g'<25$~mag).  The recovery rate heatmap (Figure~\ref{fig:heatmap_ellipse_no_nuc}) show that there is no significant difference between the performance of ellipse subtraction and denoising autoencoder when the galaxy structure is simple.

For nucleated elliptical galaxies, the performance is similar but our denoising autoencoder does not remove the light from central nuclei, because the DAE training set did not include nucleated galaxies. Therefore, the recovery rate at the very center is slightly lower for the denoising autoencoder than for the ellipse subtraction (Figure \ref{fig:heatmap_ellipse_w_nuc}), which does model and subtract the nucleus. VCC~0033 in Figure~\hyperref[fig:5a]{5(a)} is an example of such a galaxy. 

Since the DAE does not subtract the nucleus, its worse recovery rate is essentially a crowding problem. Whether one wants to subtract the nucleus, however, will depend on the science goals. If the goal is to detect all non-nuclear sources, then training the DAE with nucleated galaxies will mitigate this problem. However, if the goal is to detect a possible nucleus, then training without nuclei will help reveal them. 

In the edge-on galaxies, where the poorer fitting of ellipse subtraction leaves X-shaped residuals on the image, the advantage of DAE over ellipse model subtraction becomes more significant. The denoising autoencoder performs better on these galaxies, especially in the inner regions (Figure~\ref{fig:heatmap_edge_on}). VCC~1304 in Figure~\hyperref[fig:5b]{5(b)} is a typical example.

For smooth spiral galaxies, the ellipse fitting cannot model the spiral arms of the galaxies well, leading to both under-subtraction and over-subtraction in the images. The denoising autoencoder can remove these structures more effectively, yielding higher recover rates for faint sources near galaxy center (Figure~\ref{fig:heatmap_smooth_spiral}).  VCC 0407 in Figure~\hyperref[fig:5c]{5(c)}  is a typical example.

\begin{figure*}
    \centering

        \includegraphics[width=\linewidth]{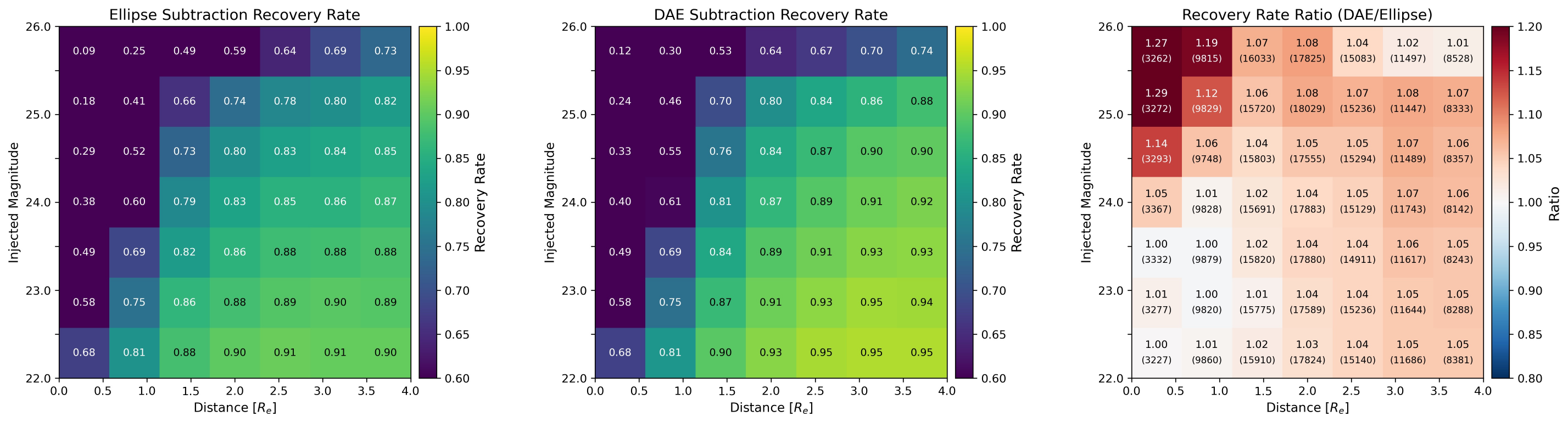}
        \label{fig:heatmap_star_forming}

    \caption{Aggregated recovery rate and mean magnitude error heatmaps created from globular clusters injected into residual images for star-forming galaxies}
    \label{fig:heatmap_bundle2}
\end{figure*}

In star-forming galaxies with irregular structures and clumpy light distributions, both methods face challenges. Recovery rates are lower compared to smoother galaxies, but the denoising autoencoder still achieves better results overall (Figure~\hyperref[fig:heatmap_star_forming]{9}). The complex structure of star-forming galaxies make ellipse fitting prone to catastrophic failure while the DAE can perform robustly (e.g., VCC 1725 in Figure~\hyperref[fig:5d]{5(d)}). 

Overall, ellipse subtraction works well on face-on elliptical galaxies with a smooth brightness profile. However, for more complex morphologies (spiral, edge-on, and star-forming galaxies), the denoising autoencoder consistently performs better than ellipse subtraction by producing cleaner residuals and a higher recovery rate for injected sources.

\mdseries

\section{Discussion and Conclusion}
\bfseries
In this paper, we present the use of a convolutional image denoising autoencoder (DAE) as a new approach to performing galaxy model subtraction. We train the DAE model on simulated NGVS galaxy images generated with \texttt{GALFIT} and demonstrate its success through tests on observed galaxy images.

The most significant advantage of the DAE is that it can operate fully automatically once the image cutouts are given. Currently, the most common technique for galaxy model subtraction is ellipse fitting, which fits elliptical isophotes to the galaxy light, and optimizes isophotal parameters iteratively to minimize the residuals. The major drawback with ellipse fitting is that it may require human intervention at different stages. First, a reasonable initial guess for isophotal parameters (center, ellipticity, and position angle) must be given to the model, or the optimization may fail catastrophically. Moreover, sometimes the fitting algorithm fails to converge when all parameters are allowed to vary freely, and visual inspection is required to fix one or more parameters in the fit. As the size of available datasets expands rapidly in the future, the manual efforts needed to process galaxy images will become overwhelming. In contrast, the DAE is not model dependent and, once trained, can process an image of any galaxy in less than $0.1$ second. 
    
Furthermore, both the visual comparison of residual images and quantitative injection-recovery tests show that the DAE can achieve similar performance as ellipse fitting on smooth, face on elliptical galaxies, and produce cleaner residual images when applied to spiral, edge-on, or star-forming galaxies. One exception is the DAE cannot fit galaxies with central bulge very well, but this can potentially solved by training the DAE on GALFIT model galaxies with two components.

The current limitation for the DAE model is its fixed $512\times 512$ input size. This fixed input size works well for galaxies with $13<g'<15$~mag at the distance of the Virgo cluster, but new models with different input size will need to be trained for larger or smaller galaxies. In future surveys, generating cutouts and centering targets can itself be nontrivial when positions and magnitudes are not known. A fully automated preprocessing workflow, including cutout generation, centering, and basic magnitude estimation, will be needed to operate at scale.

In the future, the combination of accuracy, speed, and full automation makes the DAE well suited to survey-scale pipelines. We plan to (i) add simulated galaxies with more than one component into the training set, (ii) develop an automated preprocessing stage to generate cutouts and an initial center, (iii) train models for additional input sizes to accommodate galaxies of different sizes, (iv) train separate models in other filter bands to enable multi-band analyses, and (v) extend the training set beyond NGVS images. The upcoming Rubin Observatory Legacy Survey of Space and Time (LSST) is an obvious dataset that can benefit from a DAE-based pipeline. Additionally, the growing volume of JWST and Euclid galaxy imaging will provide new high-resolution training and testing data for the DAE. The DAE model offers a practical and scalable path to robust galaxy model subtraction in wide-field surveys.

\mdseries








\section{Code Availability}

The galaxy model subtraction code developed and used in this study is publicly available at \href{https://github.com/rongrong00/galaxy-denoising-autoencoder}{GitHub} and has been archived on Zenodo \citep{liu2025_zenodo}. The code is released under the MIT License, which permits use, copying, and modification of the software provided that appropriate credit is given to the original authors.

\begin{acknowledgments}
We thank Michelle Ntampaka for suggesting the use of autoencoders for this application, and for giving suggestions on the architecture of the DAE. Based on observations obtained with MegaPrime/MegaCam, a joint project of CFHT and CEA/DAPNIA, at the Canada–France–Hawaii Telescope (CFHT), which is operated by the National Research Council (NRC) of Canada, the Institut National des Science de l'Univers of the Centre National de la Recherche Scientifique (CNRS) of France and the University of Hawaii. This work is supported in part by the Canadian Advanced Network for Astronomical Research (CANFAR), which has been made possible by funding from CANARIE under the Network-Enabled Platforms program. This research used the facilities of the Canadian Astronomy Data Centre operated by the National Research Council of Canada with the support of the Canadian Space Agency. 
\end{acknowledgments}

\facility{CFHT (MegaCam)}



\bibliography{autoencoder}{}
\bibliographystyle{aasjournal}

\end{document}